\documentclass[12pt]{article}

\usepackage[height=8.85in,width=6.45in]{geometry}

\usepackage[utf8]{inputenc}
\usepackage[T1]{fontenc}
\usepackage{amsmath}
\usepackage{amssymb}
\usepackage{mathtools}
\numberwithin{equation}{section}
\allowdisplaybreaks

\usepackage{slashed}
\usepackage{braket}
\usepackage[usenames,dvipsnames,svgnames,table]{xcolor}
\usepackage[colorlinks,citecolor=DarkGreen,linkcolor=FireBrick,urlcolor=FireBrick,linktocpage]{hyperref}
\usepackage{cite}
\usepackage{graphicx}
\usepackage{times}
\usepackage[scaled]{couriers}

\usepackage{bm}
\usepackage{subfig}

\usepackage{tikz,pgf}
\usepackage{tikz-cd}
\usetikzlibrary{shapes}
\usetikzlibrary{calc}
\usetikzlibrary{decorations.pathmorphing}
\usetikzlibrary{decorations.pathreplacing,shapes.misc}
\usetikzlibrary{positioning}
\usetikzlibrary{decorations.markings}
\tikzset{->-/.style={decoration={
  markings,
  mark=at position .5 with {\arrow{>}}},postaction={decorate}}}
\tikzset{-<-/.style={decoration={
  markings,
  mark=at position .5 with {\arrow{<}}},postaction={decorate}}}

\usepackage{xcolor}
  \definecolor{rblue}{RGB}{81, 49, 193}
  \definecolor{rorange}{RGB}{255, 147, 40}
  \definecolor{rgreen}{RGB}{176, 233, 0}

\renewcommand{\tilde}{\widetilde}

\newcommand{\Z}{\mathbb{Z}}

\renewcommand{\hat}{\widehat}

\usepackage{mdframed}

\def\Sq{\mathop{\mathrm{Sq}}\nolimits}

\begin{document}

\begin{center}

{\large \bfseries Lattice construction of exotic invertible topological phases}

\bigskip
\bigskip
\bigskip

Ryohei Kobayashi
\bigskip
\bigskip
\bigskip

\begin{tabular}{ll}
 Institute for Solid State Physics, \\
University of Tokyo, Kashiwa, Chiba 277-8583, Japan\\

\end{tabular}

\vskip 1cm

\end{center}

\noindent
In this paper, we provide state sum path integral definitions of exotic invertible topological phases proposed in the recent paper by Hsin, Ji, and Jian. The exotic phase has time reversal ($T$) symmetry, and depends on a choice of the spacetime structure called the Wu structure. The exotic phase cannot be captured by the classification of any bosonic or fermionic topological phases, and thus gives a novel class of invertible topological phases. When the $T$ symmetry defect admits a spin structure, our construction reduces to a sort of the decorated domain wall construction, in terms of a bosonic theory with $T$ symmetry defects decorated with a fermionic phase that depends on a spin structure of the $T$ symmetry defect. 
By utilizing our path integral, we propose a lattice construction for the exotic phase that generates the $\Z_8$ classification of the (3+1)d invertible phase based on the Wu structure.
This generalizes the $\Z_8$ classification of the $T$-symmetric (1+1)d topological superconductor proposed by Fidkowski and Kitaev.
On oriented spacetime, this (3+1)d invertible phase with a specific choice of Wu structure reduces to a bosonic Crane-Yetter TQFT which has a topological ordered state with a semion on its boundary.
Moreover, we propose a subclass of $G$-SPT phases based on the Wu structure labeled by a pair of cohomological data in generic spacetime dimensions. This generalizes the Gu-Wen subclass of fermionic SPT phases.

\setcounter{tocdepth}{2}
\tableofcontents

\section{Introduction}
The classification of gapped phases is an important problem in condensed matter physics. Though the classification problem is difficult and unsolved in general, one can simplify the problem by considering a simplified class of systems with a unique gapped ground state on arbitrary closed spatial manifolds. Such phases are called invertible topological phases.
In the presence of a global symmetry $G$, invertible topological phases are also called symmetry protected topological (SPT) phases protected by $G$.~\footnote{Precisely speaking, it is more common to define SPT phases as invertible phases which become trivial when we forget the global symmetry, following~\cite{Chen:2011pg}. For example, the (1+1)-dimensional topological superconductor (Kitaev wire) is not counted as an SPT phase based on such a definition, though it is an invertible topological phase. In this thesis, we do not make a careful distinction between the two concepts, namely  invertible phases and SPT phases.}

In the case of free fermions, a complete classification of SPT phases has been obtained using K-theory~\cite{Schnyder_2008, Kitaev2009free}. In the case of intrinsically interacting systems, the classification in general differs from the free phases~\cite{FidkowskiKitaev2011}, and we have to use a completely different method to perform the classification~\cite{Chen:2011pg, Gu:2012ib, Kapustin:2014dxa, Wang2017Interacting, Witten2016Fermion, Metlitski2014Interaction, Cheng2018Classification, wanggu1703, qingrui}. For example, a large class of interacting bosonic SPT phases with the global symmetry can be classified by utilizing group cohomology~\cite{Chen:2011pg}. For the case of fermionic systems, the SPT phases have a richer classification than the bosonic phases; for example, Gu and Wen found that a subclass of fermionic SPT phases is classified by a pair of cohomological data~\cite{Gu:2012ib}, generalizing the classification of bosonic phases. These fermionic SPT phases are called Gu-Wen phases or super-cohomology phases.
Later, a comprehensive classification scheme of SPT phases utilizing cobordism group is proposed in~\cite{Kapustin:2014dxa}, which is thought to classify invertible field theories which effectively describe SPT phases. The cobordism group provides a generic and powerful framework that correctly predicts the classification of interacting invertible topological phases, based on onsite (0-form) global symmetries, time reversal symmetry, or higher-form symmetries~\cite{Kapustin:2014dxa, Freed:2016rqq, Yonekura:2018ufj, Guo:2018vij, Wan_2019}.

So far, almost all the invertible topological phases discussed in literature has been either bosonic or fermionic. This implies that the effective field theories describing these topological phases require either oriented or spin structure (or these spacetime structures twisted by other symmetry groups such as pin$^+$ or pin$^-$) of the spacetime.~\footnote{Here we assume that the effective field theories are Lorentz invariant, and thereby oriented $SO(d)$ or spin group $Spin(d)$ contains the Lorentz group. Recently, there are also classes of topological phases whose effective field theories lack the Lorentz invariance, such as fractons~\cite{Haah_2011,Yoshida_2013,Vijay_2015,Vijay_2016} or those realized by foliated field theories~\cite{Shirley_2019}. These theories are beyond the scope of the present paper.} As such, the classification of invertible topological phases based on the cobordism groups has been performed for the corresponding field theories with (possibly twisted) oriented or spin structure.

Recently, it was proposed in~\cite{hsin2021exotic} that there are nontrivial invertible topological field theories based on the spacetime structure called Wu structure, which is inequivalent to any (possibly twisted) oriented or spin structure previously discussed in literature, and thus gives a new class of invertible field theories which are phrased neither as bosonic nor fermionic.
The topological phases that depend on the spacetime Wu structure are called exotic topological phases.
Wu structure in $d$ spacetime dimensions corresponds to the global symmetry given by a specific nontrivial mixture of the spacetime Lorentz symmetry $O(d)$ and the 1-form $\Z_2$ symmetry. 
Since the Lorentz group is taken as $O(d)$, the exotic topological phases possess the time reversal symmetry.
Mathematically, the symmetry is described by a specific 2-group that corresponds to a sort of the extension of $O(d)$ by the 1-form $\Z_2$ symmetry. This generalizes spin/pin structure required for fermionic systems, where one extends the Lorentz symmetry $SO(d)$ or $O(d)$ by the ordinary (0-form) $\Z_2$ symmetry that corresponds to the $\Z_2$ fermion parity. 

In this paper, we explore a lattice realization of exotic topological phases. We define  topologically invariant path integral constructions of exotic invertible topological phases in and more than (3+1)d in terms of a state sum on a lattice. In particular, the existence of an explicit state sum for the topological path integral of exotic invertible topological phases suggests that there should also exist corresponding commuting projector Hamiltonians, tensor network descriptions, and explicit quantum circuits that prepare these exotic topological phases.
We leave it to future work to explicitly develop these descriptions.

\subsection*{Summary of the main results}
Here we summarize the main results of the paper. First, we propose a state sum path integral for a specific theory $z(\eta,a)$ on a spacetime $d$-manifold $M$ equipped with a triangulation, coupled with the Wu structure $\eta$ and the $(d-3)$-form $\Z_2$ symmetry whose background gauge field is given by $a\in Z^{d-2}(M,\Z_2)$. The partition function $z(\eta,a)$ is not topologically invariant, but has an 't Hooft anomaly of the $(d-2)$-form $\Z_2$ symmetry characterized by the $(d+1)$-dimensional response theory
\begin{align}
    (-1)^{\int\Sq^3(a)}=(-1)^{\int a\cup_{d-5}a},
    \label{eq:anomalyofrefinement}
\end{align}
where $\Sq^i$ denotes the Steenrod operation of cohomology groups~\cite{MS} and $\cup_i$ is higher cup product, see Appendix~\ref{app:cup} for their review.
That is, if one considers gauge transformation and the re-triangulation of the spacetime, the partition function transforms as
\begin{align}
    z(M';\eta',a')=(-1)^{\int_{\tilde{M}}\Sq^3(a)}z(M;\eta,a),
\end{align}
where $\tilde{M}=M\times[0,1]$ is a $(d+1)$-manifold that interpolates $M$ and $M'$ equipped with the Wu structure and the $(d-2)$-form gauge field $a$, which restrict to $(\eta,a)$ and $(\eta', a')$ on $M$ and $M'$ respectively. $M'$ is a $d$-manifold which is the same as $M$ with a different triangulation and the gauge field $a'$, equivalent to $a$ in cohomology $[a]=[a']$. 

One can further show that the partition function becomes a quadratic function of the $(d-2)$-form gauge field $a\in Z^{d-2}(M,\Z_2)$ which satisfies the quadratic property
\begin{align}
    z(\eta,a)z(\eta,b)=z(\eta, a+b)(-1)^{\int_{M} a\cup_{d-4} b}.
\end{align}

The construction of the theory $z(\eta,a)$ is done by a sort of decorated domain wall construction, namely it involves decorating the symmetry defect with a topological phase in lower spacetime dimensions.
That is, $z(\eta,a)$ is given by decorating the codimension-1 defect of the time-reversal ($T$) symmetry of a bosonic phase with a specific gapped theory, which is expressed as a path integral of Grassmann variables supported on the $T$ symmetry defect. 
The Grassmann integral on the $T$ symmetry defect is essentially the same as that utilized to express the partition function of fermionic SPT phases based on the spin structure by Gu and Wen~\cite{Gu:2012ib}. 
In particular, when the $T$ symmetry defect admits the spin structure, we find that the Wu structure can be specified by a choice of spin structure on the codimension-1 $T$ symmetry defect. If the Wu structure is given in this manner, our path integral of $z(\eta,a)$ is precisely given by a decorated domain wall construction, where we decorate the $T$ symmetry defect of a bosonic topological phase with the fermionic topological phase based on the spin structure of the $T$ symmetry defect.

In the case of the spacetime dimension $d=4$, the theory $z(\eta,a)$ gives a topologically invariant path integral, since the 't Hooft anomaly~\eqref{eq:anomalyofrefinement} vanishes in $d=4$.
Actually, when the spacetime is oriented, the Wu structure can be taken to be trivial, 
so in that case the exotic topological phase reduces to a bosonic topological phase. In particular, on oriented 4-manifolds $z(\eta,a)$ reproduces a (3+1)d bosonic Crane-Yetter TQFT based on a unitary modular tensor category $\mathcal{C}=\{1,s\}$ that has a single nontrivial anyon $s$ which is a semion.

For the case of $d=4$, one can gauge the 1-form $\Z_2$ symmetry of $z(\eta,a)$, which gives the Arf-Brown-Kervaire (ABK) invariant based on the quadratic refinement $z(\eta,a)$ of $[a]\in H^2(M,\Z_2)$. The resulting theory
turns out to generate the $\Z_8$ classification of the exotic invertible topological phase~\cite{hsin2021exotic}.
This generalizes the $\Z_8$ classification of the (1+1)d pin$^-$ invertible topological phases that corresponds to the (1+1)d topological superconductor proposed by Fidkowski and Kitaev~\cite{FidkowskiKitaev2011}.

In addition, by utilizing the lattice construction of $z(\eta,a)$ we propose a subclass of the exotic invertible topological phases with an onsite (0-form) symmetry $G$, which contains anti-unitary time-reversal symmetry. In this paper, we limit ourselves to the case of $G=G_0\times \Z_2^T$ with $\Z_2^T$ a time-reversal symmetry, and the total symmetry group is direct product of $G_0$ and the 1-form symmetry of the Wu structure. It should be interesting to consider the nontrivial 2-group that involves $G$ and the 1-form $\Z_2$ symmetry of the Wu structure. We comment on the expectation about some of twisted cases in the discussion (Sec.~\ref{sec:discussion}), and leave the study on such a twisted Wu structure by $G$ for future work.

Then, we propose a subclass of exotic SPT phases labeled by a pair of the cohomological data
\begin{align}
    (\nu_d, n_{d-2})\in C_{\rho}^{d}(BG,U(1))\times Z^{d-2}(BG,\mathbb{Z}_2),
\end{align}
which are subject to the constraint
\begin{align}
    \delta_{\rho}\nu_{d}=\frac{1}{2}\mathrm{Sq}^3(n_{d-2}) \mod 1,
    \label{eq:guwengeneralize}
\end{align}
where $\rho$ denotes the twisted $G$-action on $U(1)$ that acts by complex conjugation. This generalizes the Gu-Wen phase~\cite{Gu:2012ib} that constitutes the subclass of the fermionic SPT phases, labeled by $(\nu_d, n_{d-1})\in C_{\rho}^{d}(BG,U(1))\times Z^{d-1}(BG,\mathbb{Z}_2)$ subject to the Gu-Wen equation which basically replaces $\Sq^3$ in~\eqref{eq:guwengeneralize} with $\Sq^2$.

In general, by utilizing the theory $z(\eta, a)$, one can produce a state sum path integral of an exotic topological phase, starting with a bosonic theory with a $(d-3)$-form $\Z_2$ symmetry which has a specific 't Hooft anomaly. That is, we consider a partition function of a bosonic topological phase $Z_b(a)$ coupled with a $(d-2)$-form $\Z_2$ gauge field $a\in Z^{d-2}(M,\Z_2)$. Let us assume that the bosonic theory $Z_b(a)$ has an 't Hooft anomaly characterized by the response action $(-1)^{\int\Sq^3(a)}$. Then, the partition function of the exotic invertible phases can be constructed as
\begin{align}
    Z(\eta)\propto \sum_{[a]\in H^{d-2}(M,\Z_2)} Z_b(a)z(\eta,a).
\end{align}
This construction of exotic topological phases generalizes that of fermionic topological phases by Gaiotto and Kapustin~\cite{Gaiotto:2015zta} based on fermion condensation within the path integral framework.

This paper is organized as follows. After reviewing Wu structure in Sec.~\ref{sec:wu}, we provide a path integral definition for the theory coupled with the Wu structure in Sec.~\ref{sec:zdef}. In Sec.~\ref{sec:4d}, we study the spacetime dimensions $d=4$, and construct an invertible topological phase that generates the $\Z_8$ classification. We also show that the phase is equivalent to a bosonic Crane-Yetter TQFT in the oriented case with a specific choice of the Wu structure. In Sec.~\ref{sec:guwen}, we propose a Gu-Wen type subclass of the $G$-SPT phases based on the Wu structure.

\section{Review on Wu structure}
\label{sec:wu}
In this section, we briefly review for Wu structure of the spacetime in continuum. First of all, let us recall spin structure of the spacetime required for fermionic phases. Wu structure is then regarded as a natural generalization of spin structure.

\subsection*{spin structure}
A relativistic quantum field theory in $d$ spacetime dimensions possesses the Lorentz $SO(d)$ symmetry. However, since fermions are spinors, fermions transform according to the double cover of $SO(d)$, which is $Spin(d)$. 
To define the field theory on a generic spacetime manifold, one needs to consider a $SO(d)$ bundle $\phi: M\to BSO(d)$, which is the tangent bundle $TM$ of an oriented triangulated manifold $M$. In order to have fermions, the transition functions $\phi_{ij} \in SO(d)$ defined on 1-simplices $\braket{ij}$ must be lifted to $\widetilde{\phi}_{ij} \in Spin(d)$.
Since $Spin(d)$ is the group extension 
\begin{align}
    \Z_2\to Spin(d) \to SO(d)
\end{align}
whose extension is given by $w_2\in H^2(BSO(d),\mathbb{Z}_2)$. Then, $Spin(d)$ is identified as $SO(d)\times\mathbb{Z}_2$ \textit{as a set}, so we can express $\widetilde{\phi}_{ij}$ as a pair $(\phi_{ij},\eta_{ij})\in SO(d)\times\mathbb{Z}_2$.
The nontrivial group extension is reflected in the multiplication law of $\mathbb{Z}_2$ elements $\eta_{ij}$ twisted by $w_2$. Namely, for transition functions $\widetilde{\phi}_{ij} \in Spin(d)$, we have the multiplication law
\begin{align}
\widetilde{\phi}_{01} \widetilde{\phi}_{12} = (\phi_{01} \phi_{12}, \eta_{01} + \eta_{12} + w_2(\phi_{01}, \phi_{12})).
\end{align}
Due to the cocycle condition $\widetilde{\phi}_{01} \widetilde{\phi}_{12} = \widetilde{\phi}_{02}$, we find on each 2-simplex $\braket{012}$
        \begin{align}
            \eta_{01} + \eta_{12} + \eta_{02} = w_2( \phi_{01}, \phi_{12})
            \label{eq:trivializew2}
        \end{align}
        In coordinate-free notation, this is precisely the equation $\delta \eta = w_2$.

\subsection*{Wu structure}
The Wu structure is a natural generalization of the spin structure illustrated above. The symmetry groups involved are Lorentz group $O(d)$ that includes time reversal symmetry, and a 1-form $\Z_2$ symmetry denoted as $\Z_2^{[1]}$ here. Then, the total structure of the symmetry $\mathbb{G}$ is described as a nontrivial mixture called a 2-group between $O(d)$ and $\Z_2^{[1]}$ symmetry,
\begin{align}
    \Z_2^{[1]}\to \mathbb{G}\to O(d)
\end{align}
and the 2-group structure involving $O(d)$ and $\Z_2^{[1]}$ is specified by the cohomology class $w_1w_2\in H^3(BO(d),\Z_2)$ called the Postnikov class. Analogously to the nontrivial group extension for the spin structure, a nontrivial 2-group has a distinct algebra of symmetry generators, thereby has a distinct configuration of background gauge field from that of the direct product $\Z_2^{[1]}\times O(d)$.

That is, if we denote the background gauge field of $\Z_2^{[1]}$ as $\eta\in C^2(M,\Z_2)$ and that of $O(d)$ as $\phi:M\to BO(d)$, then these background gauge fields are correlated on each 3-simplex $\braket{0123}$ as
\begin{align}
    \eta_{012}+\eta_{013}+\eta_{023}+\eta_{123}=w_1(\phi_{01})w_2(\phi_{12},\phi_{23}).
\end{align}
In coordinate-free notation, this is precisely the equation $\delta \eta = w_1w_2$.

\begin{figure}[htb]
\centering
\includegraphics{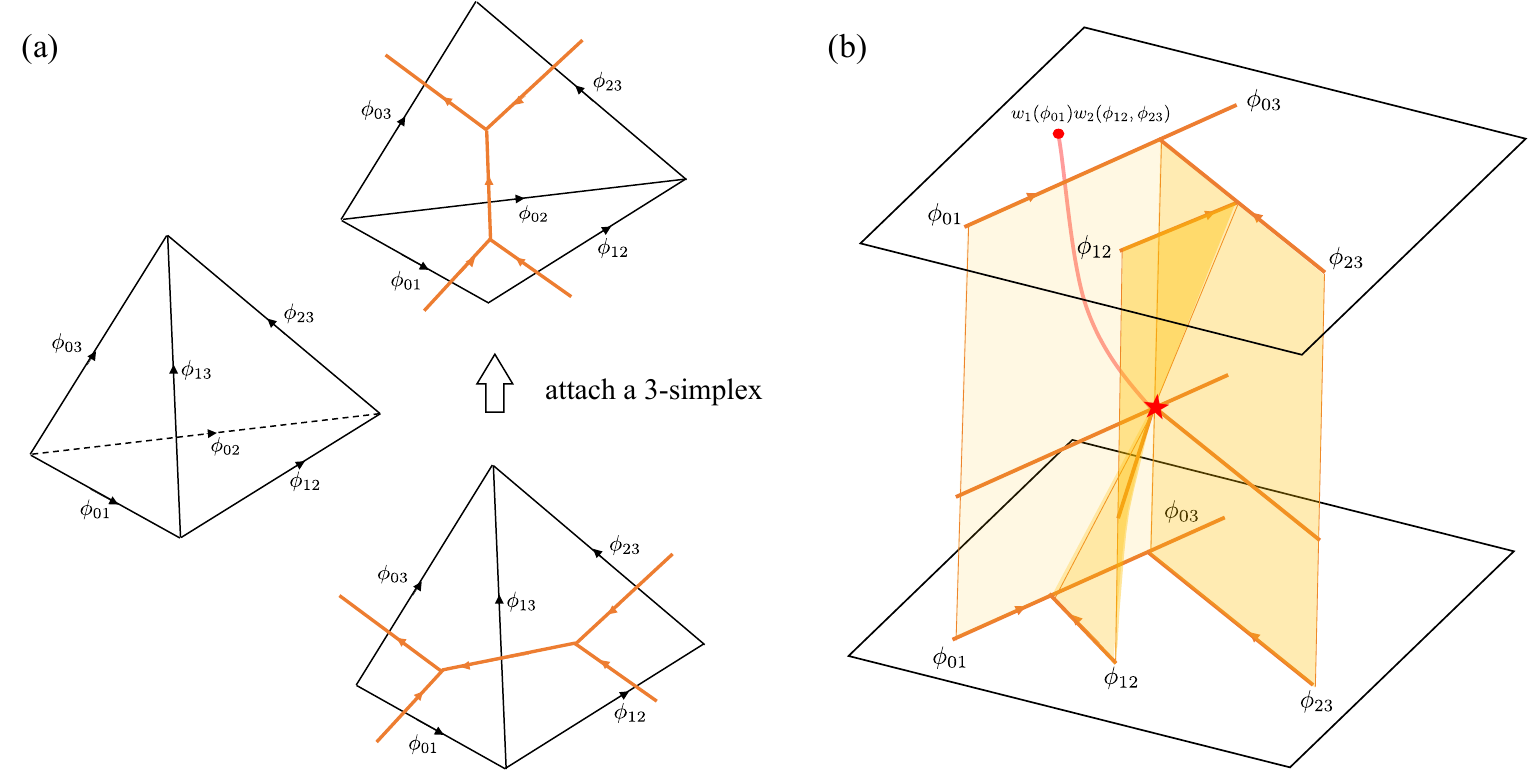}
\caption{Wu structure is the nontrivial 2-group between the Lorentz $O(d)$ symmetry and the $\Z_2^{[1]}$ 1-form symmetry. The Wu structure is specified by a choice of a $\Z_2$ 2-cochain $\eta$ with $\delta\eta=w_1w_2$, where $\eta$ gives a background gauge field of the $\Z_2^{[1]}$ symmetry. This is understood as sourcing the $\Z_2^{[1]}$ symmetry defect from the codimension-3 junction of $O(d)$ symmetry defects on each 3-simplex.}
\label{fig:2group}
\end{figure}

It is worth noting that the background $\Z_2^{[1]}$ gauge field twisted by the Postnikov class $\delta \eta = w_1w_2$ reflects the algebra for the symmetry generators of $\Z_2^{[1]}$ and $O(d)$ in the 2-group. To see this, we regard the background gauge field of $p$-form symmetry as an insertion of a $(d-p-1)$-dimensional symmetry defects on the Poincar\'e dual of the $(p+1)$-form background gauge field. Then, a 3-simplex is understood as an associator which shifts the configuration of 0-form $O(d)$ symmetry defects by the $F$-move, see Fig.~\ref{fig:2group} (a).

During the $F$-move inside a 3-simplex, the three $O(d)$ symmetry defects $\phi_{01},\phi_{12},\phi_{23}$ meets at the junction in the middle of the movie, which gives a codimension-3 junction of the $O(d)$ symmetry defects, see Fig.~\ref{fig:2group} (b). Then, the 2-group relation $\delta \eta = w_1w_2$ means that the $\Z_2^{[1]}$ symmetry generator $w_1(\phi_{01})w_2(\phi_{12},\phi_{23})$ is sourced from the codimension-3 junction. The 2-group is understood as the modification of the associativity of the 0-form $O(d)$ symmetry generators by a generator of the 1-form $\Z_2^{[1]}$ symmetry, controlled by the Postnikov class in $H^3(BO(d),\Z_2)$. See~\cite{Benini20192group, Cordova2019exploring, Tachikawa:2017gyf, kapustin2015higher} for an introduction of the 2-group.

\section{Path integral of an exotic phase}
\label{sec:zdef}
Now we construct a path integral definition for $z(\eta,a)$ on a triangulated $d$-manifold $M$ that depends on the Wu structure specified by a choice of $\eta\in C^2(M,\Z_2)$ with $\delta\eta=w_1w_2$, and a $(d-2)$-form gauge field $[a]\in H^{d-2}(M,\Z_2)$. The partition function $z(\eta,a)$ is valued in $\{\pm 1, \pm i\}$. Then, the main properties of  $z(\eta,a)$ is summarized as follows:
\begin{enumerate}
    \item The quadratic property
    \begin{align}
        z(\eta,a)z(\eta,b)=z(\eta,a+b)(-1)^{\int_M a\cup_{d-4}b}.
        \label{eq:zquad}
    \end{align}
    \item The change of $z(\eta,a)$ under the gauge transformation
$a\to a+\delta \chi$ or under the change of the triangulation 
is controlled by 't Hooft anomaly whose response action is given by 
\begin{align}
 (-1)^{\int\Sq^3 (a)}.
 \label{eq:sq3anomaly}
\end{align}
\end{enumerate}
The description of the Wu structure used for $z(\eta,a)$ is based on a specific chain $S\in Z_{d-3}(M,\Z_2)$ that represents the Poincar\'e dual of the obstruction class $w_1w_2$ for the Wu structure. Then, the Wu structure is specified by a choice of a $(d-2)$-chain $E\in C_{d-2}(M,\Z_2)$ with $\partial E=S$, which is Poincar\'e dual to $\eta\in C^2(M,\Z_2)$. As discussed in Appendix~\ref{app:w1w2}, we prepare $S$ based the choices of the cochain and chain representative of $w_1$ and $w_2$ respectively, and then take the pairing of them. In particular, when the spacetime manifold $M$ is oriented, we take the representative of $w_1$ as zero and thereby $S=0$. So, for oriented spacetime manifolds $\eta$ is closed and one can specify the Wu structure by $[\eta]\in H^2(M,\Z_2)$. As a warm-up, we start with the description of the simplest case $z(\eta,a)$ on oriented manifolds.

\subsection{Oriented case}
Here let $M$ be an oriented manifold. In that case $\eta$ is closed $\delta\eta=0$, and $z(\eta,0)$ is given by
\begin{align}
    z(\eta,a)=e^{2\pi i \int q(a)}(-1)^{\int_E a},
\end{align}
where $E\in Z_{d-2}(M,\Z_2)$ is the Poincar\'e dual of $\eta$, and 
\begin{align}
    q(a)=\frac{1}{4}(\hat{a}\cup_{d-4} \hat{a}+\hat{a}\cup_{d-3}\delta \hat{a}),
    \label{eq:pontryagingen}
\end{align}
where $\hat{a}\in C^{d-2}(M,\Z_4)$ is the $\Z_4$ lift of $a$. 
When $d=4$, the above $q(a)$ gives a cohomology operation $q: H^2(M,\Z_2)\to H^4(M,\Z_4)$ known as the Pontryagin square~\cite{pontryaginsquare, Aharony2013reading, kapustin2013topological}, and thus $z(\eta=0,a)$ is topologically invariant. However, the action $z(\eta,a)$ is not topologically invariant for $d > 4$, because
\begin{align}
    \delta(q(a))=\frac{1}{2}a\cup_{d-5}a=\frac{1}{2}\Sq^3(a) \mod 1. 
    \label{eq:coboundarysq3}
\end{align}
This demonstrates the 't Hooft anomaly of $z(\eta=0,a)$ represented in~\eqref{eq:sq3anomaly}.
This can be seen by using the Leibniz rule for higher cup product
\begin{align}
    \delta(u\cup_i v)=(-1)^{p+q-i}u\cup_{i-1}v +(-1)^{pq+p+q}v\cup_{i-1}u+\delta u\cup_i v+(-1)^p u\cup_i\delta v,
    \label{eq:leibniz}
\end{align}
with $u\in C^p, v\in C^q$ cochains.
According to the Leibniz rule we have
\begin{align}
\begin{split}
    \delta(\hat{a}\cup_{d-4} \hat{a})&=2(-1)^d\hat{a}\cup_{d-5} \hat{a}+\delta\hat{a}\cup_{d-4} \hat{a}+(-1)^d\hat{a}\cup_{d-4} \delta\hat{a},\\
    \delta(\hat{a}\cup_{d-3} \delta\hat{a})&=(-1)^d\hat{a}\cup_{d-4} \delta\hat{a}-\delta\hat{a}\cup_{d-4} \hat{a}+\delta\hat{a}\cup_{d-3}\delta\hat{a}.
    \end{split}
\end{align}
Putting these expressions into $\delta (q(a))$ shows~\eqref{eq:coboundarysq3}.
Next, let us consider a quadratic property of $z(\eta,a)$ in~\eqref{eq:zquad}. 
We have
\begin{align}
    \begin{split}
        q(a)+q( b)-q(a+ b)&=\frac{1}{4} (\hat{a}\cup_{d-4} \hat{b}+\hat{b}\cup_{d-4} \hat{a} + \hat{a}\cup_{d-3}\delta\hat{b} + \hat{b}\cup_{d-3}\delta\hat{a}).
    \end{split}
\end{align}
Here, the rhs can be rewritten as
\begin{align}
    q(a)+q( b)-q(a+ b)=\frac{1}{2}{a}\cup_{d-4} {b} + \frac{1}{4}\delta(\hat{a}\cup_{d-3}\hat{b}+\delta \hat{a}\cup_{d-2} \hat{b}) \mod 1.
    \label{eq:quadofq}
\end{align}
To see this, we note that
\begin{align}
    \begin{split}
        \frac{1}{4}\delta(\hat{a}\cup_{d-3}\hat{b})&=\frac{1}{4}(\delta \hat{a}\cup_{d-3} \hat{b}+(-1)^d\hat{a}\cup_{d-3}\delta \hat{b} +(-1)^{d-1} \hat{a}\cup_{d-4} \hat{b}+(-1)^d\hat{b}\cup_{d-4} \hat{a}) \\
        &= \frac{1}{4}((-1)^{d-1}\hat{b}\cup_{d-3}\delta \hat{a} +(-1)^{d-1} \delta(\delta \hat{a}\cup_{d-2} \hat{b}) - \delta \hat{a}\cup_{d-2}\delta\hat{b} \\ & \quad + (-1)^d\hat{a}\cup_{d-3}\delta \hat{b} +(-1)^{d-1} \hat{a}\cup_{d-4} \hat{b}+(-1)^d\hat{b}\cup_{d-4} \hat{a}).
    \end{split}
\end{align}
Here we again used the Leibniz rule~\eqref{eq:leibniz}.
Since $\frac{1}{4}\delta \hat{a}\cup_{d-2}\delta\hat{b}=0 \mod 1$, we have
\begin{align}
\begin{split} 
\frac{1}{4}\delta(\hat{a}\cup_{d-3}\hat{b}) &= \frac{1}{4}((-1)^{d-1}\hat{b}\cup_{d-3}\delta \hat{a} +(-1)^{d-1} \delta(\delta \hat{a}\cup_{d-2} \hat{b})  \\ & \quad + (-1)^d\hat{a}\cup_{d-3}\delta \hat{b} +(-1)^{d-1} \hat{a}\cup_{d-4} \hat{b}+(-1)^d\hat{b}\cup_{d-4} \hat{a}) \mod 1.
    \end{split}
\end{align}
Hence,
\begin{align}
\begin{split}
    \frac{1}{2}&(\hat{a}\cup_{d-4} \hat{b}) + \frac{1}{4}\delta(\hat{a}\cup_{d-3}\hat{b}+\delta \hat{a}\cup_{d-2} \hat{b})\\&=\frac{(-1)^d}{4}(\hat{a}\cup_{d-4} \hat{b}+\hat{b}\cup_{d-4} \hat{a} + \hat{a}\cup_{d-3}\delta\hat{b} + \hat{b}\cup_{d-3}\delta\hat{a}) \mod 1.
    \end{split}
\end{align}
Thus we obtain~\eqref{eq:quadofq}.
When the spacetime manifold $M$ is oriented, only the first term of the rhs of~\eqref{eq:quadofq} contributes mod 1 to the integral on $M$. Thus we obtain the quadratic property as
\begin{align}
z(\eta,a)z(\eta,b)=z(\eta,a+b)(-1)^{\int_M a\cup_{d-4}b}.
\end{align}
This is what we want.

\subsection{Unoriented case}
\label{subsec:unoriented}
Now let us construct the Grassmann integral $z(\eta, a)$ on a $d$-manifold $M$ which might be unoriented. 
We construct an unoriented manifold by picking locally oriented patches, and then gluing them along codimension one loci by transition functions. The locus where the transition functions are orientation reversing, constitutes a representative of the dual of first Stiefel-Whitney class $w_1$. We will sometimes call the locus an orientation reversing wall, represented as a $(d-1)$-cycle $W$ Poincar\'e dual to $w_1$.
For convenience, we endow $M$ with a barycentric subdivision for the triangulation of $M$.
Namely, each $d$-simplex in the initial triangulation of $M$ is subdivided into $(d+1)!$ simplices, whose vertices are barycenters of the subsets of vertices in the $d$-simplex.
We then assign a local ordering to vertices of the barycentric subdivision, such that a vertex on the barycenter of $i$ vertices is labeled as $i$.
Each simplex can then be either a $+$ simplex or a $-$ simplex, depending on whether the ordering agrees with the orientation or not. 

\begin{figure}[htb]
\centering
\includegraphics{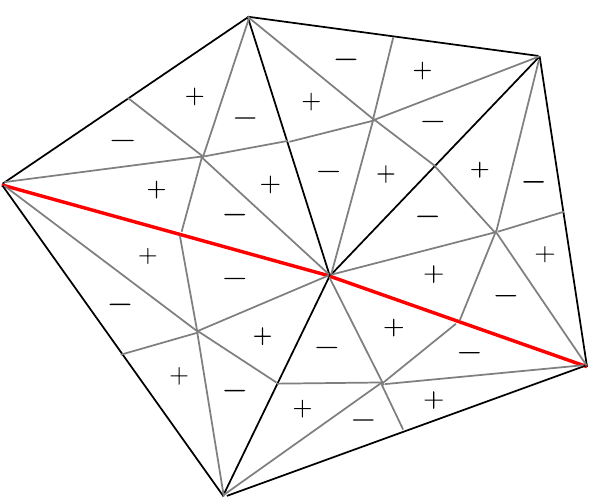}
\caption{The $\pm$ signs of $d$-simplices near the orientation reversing wall $W$ for $d=2$, which is represented as a red line. The $d$-simplices are barycentric-subdivided.}
\label{fig:wall}
\end{figure}

Then, one can explicitly obtain a $(d-3)$-cycle $S\in Z_{d-3}(M,\Z_2)$ that represents the Poincar\'e dual of $w_1w_2$ based on the barycentric subdivision. That is, $S$ is given by the set of all $(d-3)$-simplices of $W$. Since it is known that the set of all $(D-i)$-simplices in a $D$-dimensional closed manifold gives the representative of $w_i$~\cite{HalperinToledo,BlantonMcCrory}, one can see that $S$ also gives the representative for the Poincar\'e dual of $w_2$ in $W$. The proof that $S$ represents the Poincar\'e dual of $w_1w_2$ of $M$ is given in Appendix~\ref{app:w1w2}.

Analogously to what we did in the oriented case, we would like to define $z(\eta,a)$ in terms of the action $e^{2\pi i q(a)}$ with $q(a)$ defined in~\eqref{eq:pontryagingen}. However, the action $e^{2\pi i q(a)}$ does not correctly produce the quadratic property~\eqref{eq:zquad};
according to the quadratic property of $q(a)$ obtained in~\eqref{eq:quadofq}, we instead have
\begin{align}
    e^{2\pi i\int q(a)}e^{2\pi i\int q(b)}=e^{2\pi i\int q(a+b)}\cdot(-1)^{\int_{M} a\cup_{d-4} b}(-1)^{\int_{W} a\cup_{d-3} b}.
    \label{eq:qquadunoriented}
\end{align}
This is because the integration of the coboundary term $\frac{1}{4}\int\delta(\hat{a}\cup_{d-3}\hat{b}+\delta \hat{a}\cup_{d-2} \hat{b})$ in~\eqref{eq:quadofq} evaluates non-trivially on the orientation-reversing wall, as $\frac{1}{2}\int_{W}a\cup_{d-3} b \mod 1$. Hence, in order to obtain the desired quadratic property of $z(\eta,a)$, we introduce an additional term $\sigma (W;a)$ on the orientation-reversing wall, and define
\begin{align}
    z(\eta,a)=e^{2\pi iq(a)}\sigma (W;a)(-1)^{\int_E a},
    \label{eq:zunoriented}
\end{align}
where $\sigma (W;a)$ is a $(d-1)$-dimensional action supported on $W$, and $E\in C_{d-2}(M,\Z_2)$ is the Wu structure that satisfies $\partial E=S$.
If $W$ admits a spin structure, $E$ can be taken as the element of $C_{d-2}(W,\Z_2)$ since $S$ represents the Poincar\'e dual of $w_2$ on $W$. Then the Wu structure of $M$ is specified by the spin structure of $W$. In that case the action~\eqref{eq:zunoriented} is regarded as a decorated domain wall construction, where one decorates the $T$ symmetry defect of a bosonic theory with a spin (fermionic) theory $\sigma (W;a)(-1)^{\int_E a}$.

$\sigma (W;a)$ has the two main properties summarized as follows:
\begin{enumerate}
    \item The quadratic property
    \begin{align}
        \sigma (W;a)\sigma (W;a)=\sigma (W;a+b)(-1)^{\int_W a\cup_{d-3}b}.
        \label{eq:sigmaquad}
    \end{align}
    \item if one considers gauge transformation and the re-triangulation of the spacetime, the partition function transforms as
\begin{align}
 \sigma (W';a')=(-1)^{\int_{\tilde{W}}\Sq^2 (a)}(-1)^{\int_{S_{\tilde{W}}}a}\sigma (W;a).
 \label{eq:sq2anomaly}
\end{align}
where $\tilde{W}=W\times[0,1]$ is a $d$-manifold that interpolates $W$ and $W'$ equipped with the $(d-2)$-form gauge field $a$, which restrict to $a$ and $a'$ on $W$ and $W'$ respectively. $W'$ is a $(d-1)$-manifold which is the same as $W$ with a different triangulation and the gauge field $a'$, equivalent to $a$ in cohomology $[a]=[a']$. $S_{\tilde{W}}$ is a set of all $(d-2)$-simplices of $\tilde{W}$.
\end{enumerate}
Based on the quadratic property of $\sigma (W;a)$, one can immediately check that the combined action~\eqref{eq:zunoriented} satisfies the desired quadratic property~\eqref{eq:zquad}, since the extra factor $(-1)^{\int_{W} a\cup_{d-3} b}$ for the quadratic property of $e^{2\pi i q(a)}$ in~\eqref{eq:qquadunoriented} is precisely canceled by the quadratic property of $\sigma (W;a)$ in~\eqref{eq:sigmaquad}.

One can further check the invariance of $z(\eta,a)$ under moving the orientation-reversing wall $W$, which guarantees the invariance of the path integral under shifting the $T$ symmetry defect.
To see this, suppose we initially have the orientation-reversing wall $W$ which is moved to the final configuration $W'$. Let $\tilde{W}$ be a $d$-manifold that interpolates $W$ and $W'$, $\partial \tilde{W}=W\sqcup \overline{W}'$. According to the property of the Grassmann integral~\eqref{eq:sq2anomaly}, we have
\begin{align}
    \sigma (W';a|_{W'})=(-1)^{\int_{\tilde{W}}\Sq^2 (a)}(-1)^{\int_{S_{\tilde{W}}}a}\sigma (W;a|_{W}).
\end{align}
Meanwhile, since the action $e^{2\pi i q(a)}$ is complex conjugated under the orientation reversal, which means when the orientation is reversed on a single $d$-simplex
\begin{align}
    e^{2\pi i q(a)}\to e^{-2\pi i q(a)}= e^{2\pi i q(a)}\cdot(-1)^{a\cup_{d-4} a},
\end{align}
so the action $e^{2\pi i\int q(a)}$ gets shifted by $(-1)^{\int_{\tilde{W}}a\cup_{d-4} a}$.
In addition, the factor $(-1)^{\int_E a}$ gets shifted by $(-1)^{\int_{S_{\tilde{W}}}a}$. This is shown by noting that $\partial S_{\tilde{W}}=S_{W}+S_{W'}$, where $S_{W}$ (resp.~$S_{W'}$) is the set of all $(d-3)$-simplices of $W$ (resp.~$W'$), see Appendix~\ref{app:bulkboundary} for the derivation. This means that $\partial(S_{\tilde{W}}+E+E')=0$, with $E$ (resp.~$E'$) a choice of the Wu structure that satisfies $\partial E=S_{W}$ (resp.~$\partial E'=S_{W'}$). This shows that
\begin{align}
    (-1)^{\int_E a}=(-1)^{\int_{S_{\tilde{W}}}a}(-1)^{\int_{E'} a}.
\end{align}
Thus, one can see that the combined action $ z(\eta,a)$ in~\eqref{eq:zunoriented} is completely invariant under the shift of the $T$ defect, since the variation of each term is precisely canceled with each other.

This shows that the 't Hooft anomaly of $z(\eta,a)$ is the same as the oriented case, since one can freely move the $T$ symmetry defect without shifting the partition function. Thus, the 't Hooft anomaly is given by the response action~\eqref{eq:sq3anomaly},
\begin{align}
 (-1)^{\int\Sq^3 (a)}.
\end{align}

\subsubsection*{The definition of $\sigma (W;a)$: Gu-Wen Grassmann integral}
Now we provide the definition of $\sigma (W;a)$, and demonstrate the properties~\eqref{eq:sigmaquad} and~\eqref{eq:sq2anomaly}. $\sigma (W;a)$ is realized by a theory called the Gu-Wen Grassmann integral~\cite{Gu:2012ib,Gaiotto:2015zta}, expressed as a path integral of Grassmann variables supported on $W$. We note that $W$ is oriented, and the sign of each $(d-1)$-simplex of $W$ is defined by taking the sign of the neighboring $d$-simplex of $M$.

To construct the Grassmann integral~$\sigma (W;a)$, we assign a pair of Grassmann variables $\theta_e, \overline{\theta}_e$ on each $(d-2)$-simplex $e$ of $W$ such that $a(e)=1$, we associate $\theta_e$ on one side of $e$ contained in one of $(d-1)$-simplices of $W$ neighboring $e$ (which will be specified later), $\overline{\theta}_e$ on the other side.
Then, we define $\sigma (W;a)$ as
\begin{equation}
    \sigma (W;a)=\int\prod_{e|a(e)=1}d\theta_e d\overline{\theta}_e \prod_t u(t),
    \label{sigmadef}
\end{equation}
where $t$ denotes a $(d-1)$-simplex of $W$, and $u(t)$ is the product of Grassmann variables contained in $t$.
For instance, for $(d-1)=2$, $u(t)$ on $t=(012)$ is the product of
$\vartheta_{12}^{a(12)}, \vartheta_{01}^{a(01)}, \vartheta_{02}^{a(02)}$. 
Here, $\vartheta$ denotes $\theta$ or $\overline{\theta}$ depending on the choice of the assigning rule, which will be introduced later. The order of Grassmann variables in $u(t)$ will also be defined shortly.
We note that $u(t)$ is ensured to be Grassmann-even since $a$ is closed. 

\begin{figure}[htb]
\centering
\includegraphics{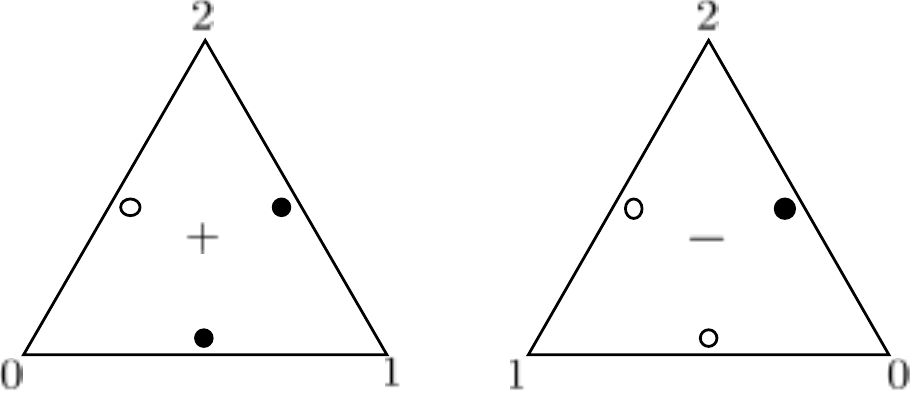}
\caption{Assignment of Grassmann variables on 1-simplices in the case of $d=2$. $\theta$ (resp.~$\overline{\theta}$) is represented as a black (resp.~white) dot.}
\label{fig:Grassmann}
\end{figure}

Due to the fermionic sign of Grassmann variables, $\sigma (W;a)$ becomes a quadratic function, whose quadratic property depends on the order of Grassmann variables in $u(t)$. We will adopt the order used in Gaiotto-Kapustin~\cite{Gaiotto:2015zta}, which is defined as follows. 
\begin{itemize}
\item
For $t=(01\dots d-1)$, we label a $(d-2)$-simplex $(01\dots\hat{i}\dots d-1)$ (i.e., a $(d-1)$-simplex given by omitting a vertex $i$) simply as $\hat{i}$. 
\item Then, the order of $\vartheta_i$ for $+$ $(d-1)$-simplex $t$ is defined by first assigning even $(d-2)$-simplices in ascending order, then odd simplices in ascending order again:
\begin{equation}
    \hat{0}\to \hat{2}\to \hat{4}\to\dots \to \hat{1}\to \hat{3}\to \hat{5}\to\dots
\end{equation}
\item For $-$ $(d-1)$-simplices, the order is defined in opposite way:
\begin{equation}
    \dots\to \hat{5}\to \hat{3}\to \hat{1} \to \dots \to \hat{4}\to \hat{2}\to \hat{0}.
\end{equation}
\end{itemize}
For example, for $(d-1)=2$, $u(012)=\vartheta_{12}^{a(12)}\vartheta_{01}^{a(01)}\vartheta_{02}^{a(02)}$ when $(012)$ is a $+$ triangle, 
and $u(012)=\vartheta_{02}^{a(02)}\vartheta_{01}^{a(01)}\vartheta_{12}^{a(12)}$ for a $-$ triangle. 
Then, we choose the assignment of $\theta$ and $\overline{\theta}$ on each $e$ such that, if $t$ is a $+$ (resp.~$-$) simplex, $u(t)$ includes $\overline{\theta}_e$ when $e$ is labeled by an odd (resp.~even) number, see Fig.~\ref{fig:Grassmann}.
Based on the above definition of $u(t)$, the quadratic property of $u(t)$ is given by~\eqref{eq:sigmaquad},
\begin{equation}
    \sigma (W;a)\sigma (W;a')=\sigma (W;a+a')(-1)^{\int_W a\cup_{d-3}a'},
    \label{eq:ClosedQuad}
\end{equation}
for closed $a, a'$. To see this, we just have to bring the product of two Grassmann integrals
\begin{equation}
    \sigma (W;a)\sigma (W;a')=\int\prod_{e|a(e)=1}d\theta_e d\overline{\theta}_e\prod_{e|a'(e)=1}d\theta_e d\overline{\theta}_e \prod_t u(t)[a]\prod_t u(t)[a']
\end{equation}
into the form of $\sigma (W;a+a')$ by permuting Grassmann variables, and count the net fermionic sign.
First of all, each path integral measure on $e$ picks up a
sign $(-1)^{a(e)a'(e)}$ by permuting $d\overline{\theta}_e^{a(e)}$ and $d\theta_e^{a'(e)}$.
For integrands, $u(t)$ on different $(d-1)$-simplices commute with each other for closed $a$, so nontrivial signs occur only by reordering $u(t)[a]u(t)[a']$ to $u(t)[a+a']$ on a single $(d-1)$-simplex. The sign on $t$ is explicitly written as
\begin{equation}
    (-1)^{\sum_{e,e'\in t}^{e>e'}a(e)a'(e')},
\end{equation}
where the order $e>e'$ is determined by $u(t)$. Hence, the net fermionic sign is given by
\begin{equation}
    \sigma (W;a)\sigma (W;a')=\sigma (W;a+a')\prod_t(-1)^{\epsilon[t, a, a']},
    \label{eq:orientedquad}
\end{equation}
with
\begin{equation}
   \epsilon[t, a, a']=\sum_{e,e'\in t, e>e'}a(e)a'(e')+\sum_{e\in t, e>0}a(e)a'(e),
   \label{eq:redis}
\end{equation}
where $e>0$ if $u[t]$ includes a $\overline{\theta}_e$ variable. The sign $\epsilon[t, a, a']$ turns out to have a neat expression in terms of the higher cup product.

At a $+$ simplex, after some efforts we can rewrite $\epsilon[t, a, a']$ as
\begin{equation}
\begin{split}
    \epsilon[t, a, a']&=\sum_{i}a_{2i+1}\cdot \delta a'(t)+\sum_{i<j}a_{2i+1}a'_{2j+1}+\sum_{i>j}a_{2i}a'_{2j}\\
    &=a\cup_{d-3}a'+a\cup_{d-2}\delta a'.
    \end{split}
    \label{eq:pepsilon}
\end{equation}

At a $-$ simplex, similarly we have
\begin{equation}
\begin{split}
    \epsilon[t, a, a']&=\sum_{i}a_{2i}\cdot \delta a'(t)+\sum_{i<j}a_{2i+1}a'_{2j+1}+\sum_{i>j}a_{2i}a'_{2j}\\
    &=\delta a(t)\delta a'(t)+a\cup_{d-3}a'+a\cup_{d-2}\delta a'.
    \end{split}
    \label{eq:mepsilon}
\end{equation}
We can see the quadratic property~\eqref{eq:ClosedQuad} when $a, a'$ are closed. 

When $a=\delta\lambda$ for some $\lambda\in C^{d-3}(W,\Z_2)$, the Grassmann integral can be explicitly computed as
\begin{align}
    \sigma (W;\delta\lambda)=(-1)^{\int_W\lambda\cup_{d-4}\delta\lambda+\lambda\cup_{d-5}\lambda}(-1)^{\int_{S}\lambda},
    \label{eq:sigmacoboundary}
\end{align}
see~\cite{Gaiotto:2015zta} for its derivation. 
Now let us show the effect of gauge transformation and re-triangulation of $\sigma (W;a)$ given in~\eqref{eq:sq2anomaly}. To see this, let us introduce an expression of $\sigma (W;a)$ convenient for our purpose. Let us assume that $W$ equipped with the background gauge field $a\in Z^{d-2}(W,\Z_2)$ is null-bordant, i.e., $W$ is a boundary of some oriented $d$-manifold $X$ and $a$ is extended to $X$. Then, one can consider the Wess-Zumino-Witten (WZW) like expression of the Grassmann integral
\begin{align}
    \sigma (W;a)= (-1)^{\int_X\Sq^2 a}(-1)^{\sum_{S_X}a},
    \label{eq:WZW}
\end{align}
where $S_X$ is a set of all $(d-2)$-simplices of $X$. 
Due to the Wu relation~\cite{ManifoldAtlasWu}, $\Sq^2(a)+w_2\cup a$ is exact for an arbitrary oriented $d$-manifold. Hence, the above expression does not depend on the extending manifold $X$.
We can explicitly check that~\eqref{eq:WZW} satisfies the properties of the Grassmann integral~\eqref{eq:ClosedQuad},~\eqref{eq:sigmacoboundary}. First, let us check the quadratic property of the WZW-like expression,
\begin{align}
\begin{split}
    \sigma (W;a)\sigma (W;a')&=\sigma (W;a+a')(-1)^{\int_{X}(a\cup_{d-4}a'+a'\cup_{d-4}a)} \\
    &= \sigma (W;a+a')(-1)^{\int_W a\cup_{d-3}a'}.
    \end{split}
\end{align}
Next, when $a=\delta\lambda$ for some $\lambda\in C^{d-3}(X,\Z_2)$, we have
\begin{align}
\begin{split}
    \sigma (W;\delta\lambda)&=(-1)^{\int_X\Sq^2\delta\lambda}(-1)^{\sum_{S_X}\delta\lambda} \\
    &= (-1)^{\int_W \lambda\cup_{d-4}\delta\lambda+\lambda\cup_{d-5}\lambda}(-1)^{\sum_{S_X}\lambda},
    \end{split}
\end{align}
where we used $\partial S_X=S$, namely the boundary of $S_X$ gives the dual of $w_2$ on $W$.
See Appendix~\ref{app:bulkboundary} for the derivation of $\partial S_X=S$.

Since the WZW-like expression satisfies the key properties~\eqref{eq:ClosedQuad},~\eqref{eq:sigmacoboundary}, one can identify the original definition of $\sigma (W;a)$ as the WZW-like expression up to gauge invariant counterterms, i.e., the Grassmann integral is expressed in the form of
\begin{align}
    \sigma (W;a)=(-1)^{\int_X\Sq^2 a}(-1)^{\sum_{S_X}a}\epsilon(a),
\end{align}
where $\epsilon: H^{d-2}(W,\Z_2)\to\{\pm 1\}$ is gauge invariant. 
The additional term $\epsilon(a)$ does not affect on the response to gauge transformation or re-triangulation, so we can identify $\sigma (W;a)$ as the WZW-like expression for a practical purpose.

Based on the WZW-like expression, we immediately know the effect of re-triangulation as follows.
Suppose we have two configurations of $a$ and triangulations on $W\times\{0\}$ and $W\times\{1\}$ interpolated by $\tilde{W}=W\times[0,1]$. Then, according to the WZW-like expression for ${\sigma}_{d-1}(W\times\{0\})\sigma (W\times\{1\})$, up to gauge invariant counterterms $\sigma (W\times\{0\})$ is given by
\begin{equation}
    \sigma (W\times\{0\})=(-1)^{\int_{\tilde{W}}\Sq^2a}(-1)^{\sum_{S_{\tilde{W}}}a}\cdot\sigma (W\times\{1\}),
    \label{eq:pinretriangulate}
\end{equation}
where $a$ on $W\times\{0\}$, $W\times\{1\}$ is extended to $\tilde{W}$. This expression directly shows that the effect of gauge transformation and re-triangulation of $\sigma(W,a)$ is controlled by the bulk response action
\begin{align}
    (-1)^{\int_W\Sq^2a}(-1)^{\sum_{S_{\tilde{W}}}a}.
\end{align}
This proves~\eqref{eq:sq2anomaly}.

\section{(3+1)d exotic invertible phase: $\Z_8$ classification}
\label{sec:4d}

   \subsection{Arf-Brown-Kervaire invariant: Kitaev type phase}
   In the spacetime dimensions $d=4$, the partition function $z(\eta,a)$ is free of 't Hooft anomaly and gives a topologically invariant path integral, with a quadratic property for $[a]\in H^2(M,\Z_2)$ given by
   \begin{align}
        z(\eta,a)z(\eta,b)=z(\eta,a+b)(-1)^{\int_M a\cup b}.
        \label{eq:zquad4}
    \end{align}
Then the partition function for the exotic invertible phase is obtained by gauging the $\Z_2$ 1-form symmetry
\begin{align}
    Z_{\mathrm{ABK}}(\eta)=\frac{1}{\sqrt{|H^2(M,\mathbb{Z}_2)|}}\sum_{a\in H^2(M,\mathbb{Z}_2)}z(\eta,a).
    \label{eq:ABK}
\end{align}
This is the Arf-Brown-Kervaire (ABK) invariant based on the quadratic function $z(\eta,a)$, valued in 8th root of unity.
The theory generates the $\Z_8$ classification of the exotic invertible topological phase proposed in~\cite{hsin2021exotic}.
This generalizes the $\Z_8$ classification of the (1+1)d pin$^-$ invertible topological phases that corresponds to the (1+1)d topological superconductor~\cite{Kitaev00unpaired} proposed by Fidkowski and Kitaev~\cite{FidkowskiKitaev2011}, whose partition function is given by the ABK invariant based on the quadratic refinement of $H^1(M,\Z_2)$, see~\cite{KirbyTaylor,Thorngren2018bosonization, Kobayashi2019pin}.

\subsection{Relation to the Crane-Yetter model in oriented case}
When the spacetime 4-manifold is oriented, the above theory $Z_{\mathrm{ABK}}(\eta)$ with a specific choice of the Wu structure $\eta=0$
is identical to the Crane-Yetter TQFT~\cite{crane1993, WalkerWang2011} based on a unitary modular tensor category (UMTC) whose objects are $\{1,s\}$, with $s$ a semion with $s\times s=1, \theta_s=i$. 
The above ABK invariant on oriented spacetime 4-manifolds has been discussed in~\cite{Bhardwaj2020SL2Z, morita}, and known to be the signature $\sigma$ of the oriented 4-manifold mod 8 when the Wu structure is chosen as $\eta=0$,
\begin{align}
Z_{\mathrm{ABK}}(\eta=0)=\exp\left(2\pi i\cdot \frac{\sigma}{8}\right).
\end{align}
Meanwhile, the partition function of the Crane-Yetter TQFT is also a signature,
\begin{align}
Z_{\mathrm{CY}}=\exp\left(2\pi ic_-\cdot \frac{\sigma}{8}\right),
\end{align}
where $c_-$ is the topological central charge (i.e., framing anomaly) mod 8 of the given UMTC. Since we have $c_-=1$ mod 8 for the UMTC $\{1,s\}$, the partition function is identical $Z_{\mathrm{ABK}}(\eta=0)=Z_{\mathrm{CY}}$.
In the rest of this section, let us demonstrate the relation between the above invariant $Z_{\mathrm{ABK}}(\eta)$ in~\eqref{eq:ABK} and the Crane-Yetter TQFT at the level of the microscopic action.

The Crane-Yetter TQFT takes as input a UMTC $\mathcal{C}$ describing the (2+1)d topological order. It associates a partition function $Z_{\mathrm{CY}}(M)$ to a triangulated 4-manifold $M$ with branching structure. See~\cite{Kitaevanyons, Bonderson07b} for a review of UMTC.

The state sum consists of a summation over all possible assignments of the following data:
        \begin{itemize}
            \item To each 2-simplex $(ijk)$, assign a simple object (anyon) $a_{ijk}  \in \mathcal{C}$
            \item To each 3-simplex $\braket{ijkl}$, assign an anyon $b_{ijkl} \in \mathcal{C}$ and an element of the fusion and splitting space $V^{b}_{ijl,jkl} \otimes V_b^{ikl,{ijk}}$ .
        \end{itemize}
        
\begin{figure}[htb]
           \centering
           \includegraphics{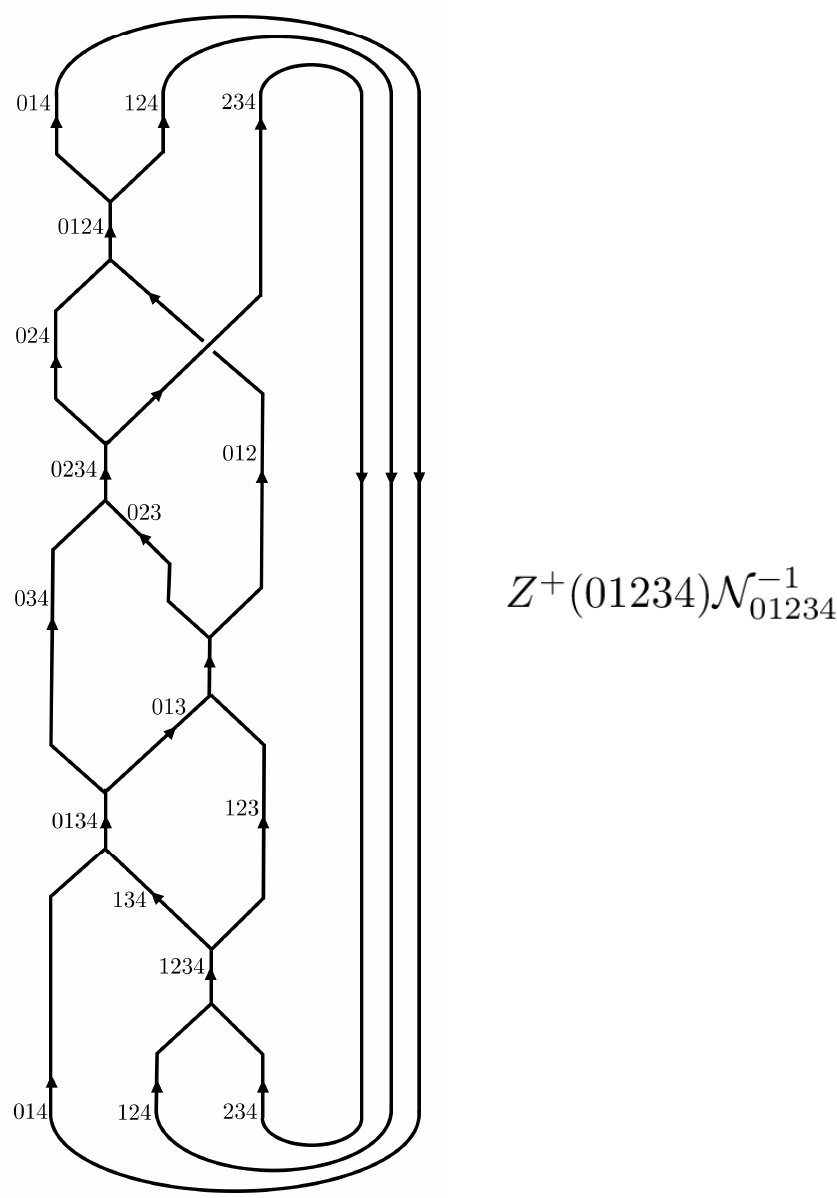}
           \caption{The amplitude associated to a $+$ 4-simplex 01234, where the branching structure induces the order on the labels. Black lines are anyon lines taking values in the input UMTC $\mathcal{C}$.
           }
           \label{fig:15jSymbols}
       \end{figure}

For convenience, we will often ignore the distinction between a simplex and the anyon or group element data assigned to it, i.e. simply write $a_{ijk}$ as $ijk$, or $b_{ijkl}$ as $ijkl$. 
        We then assign an amplitude $Z_b^{\epsilon(\Delta_4)}(\Delta_4)$ to each 4-simplex $\Delta_4$ of $M$, with $\epsilon(\Delta_4)$ a sign of the $4$-simplex determined by the branching structure. This amplitude is given diagrammatically in Fig.~\ref{fig:15jSymbols}, with a normalization factor
       \begin{equation} \label{eq:normalizationFactor_15j}
           \mathcal{N}_{01234} = \sqrt{\frac{\prod_{\Delta_3 \in \text{3-simplices}} d_{b_{\Delta_3}}}{\prod_{\Delta_2 \in \text{2-simplices}} d_{a_{\Delta_2}}}} .
       \end{equation}
The 15j symbol is explicitly given by
\begin{align}
Z^+(01234) =  &\sum_{d,a \in \mathcal{C}}  F^{024,234,012}_{d,0234,a} R_a^{012,234} (F^{024,012,234}_{d})^{-1}_{a,0124}F^{014,124,234}_{d,0124,1234}\times \nonumber \\
&(F^{014,134,123}_{d})^{-1}_{1234,0134}F^{034,013,123}_{d,0134,0123} \times (F^{034,023,012}_{d})^{-1}_{0123,0234}
\label{eqn:Zplusdef}
\\
Z^-(01234) =  &\sum_{d,a \in \mathcal{C}} (F^{024,234,012}_{d})^{-1}_{a,0234} (R_a^{012,234})^{-1} F^{024,012,234}_{d,0124,a}\left(F^{014,124,234}_{d}\right)^{-1}_{1234,0124}\times \nonumber\\
& F^{014,134,123}_{d,0134,1234} \left(F^{034,013,123}_d\right)^{-1}_{0123,0134}\times F^{034,023,012}_{d,0234,0123}
\label{eqn:Zminusdef}
\end{align}
on a $+$ and $-$ 4-simplex, respectively.
Then we can define the full path integral of the Crane-Yetter TQFT. Let $\chi$ be the Euler characteristic of $M$, and let $T^k$ be the set of $k$-simplices of $M$. Then define 
       \begin{equation}
           Z_{\mathrm{CY}}(M)=\mathcal{D}^{2(N_0-N_1)-\chi}\sum_{\lbrace a,b \rbrace} \frac{\prod_{\Delta_2\in T^2} d_{a_{\Delta_2}} \prod_{\Delta_4 \in T^4}Z^{\epsilon(\Delta_4)}(\Delta_4)}{\prod_{\Delta_3 \in T^3} d_{b_{\Delta_3}}}
       \end{equation}
       where $N_k$ is the number of $k$-simplices of $M$, $d_a$ is quantum dimension of an anyon $a$, and $\mathcal{D}=\sqrt{\sum_{a\in\mathcal{C}}d_a^2}$ is the total dimension.
       
Now let us demonstrate the equivalence of the Crane-Yetter path integral to the ABK invariant in~\eqref{eq:ABK} for the case that $\mathcal{C}$ is a semion theory $\mathcal{C}=\{1,s\}$. 
The nontrivial $F$ and $R$ symbols of $\mathcal{C}=\{1,s\}$ are 
$F^{sss}_{s}=-1, R^{ss}_1=i$, otherwise $1$~\cite{rowell2009classification}.

Since the $F^{abc}_{d}$ is real and symmetric under the permutation of labels $a,b,c$ for our UMTC, we can see that
\begin{align}
Z^+(01234) =  & R_a^{012,234} F^{014,124,234}_{d} F^{014,134,123}_{d}F^{034,013,123}_{d}F^{034,023,012}_{d},
\label{eq:15jsimple}
\end{align}
and we have $Z^-(01234)=(Z^+(01234))^{-1}$. $a,d\in \mathcal{C}$ that appears in~\eqref{eq:15jsimple} is uniquely determined since $\mathcal{C}=\{1,s\}$ is abelian.

Now, let us consider a 2-form $\Z_2$ field $a\in Z^2(M,\Z_2)$ and we take the anyon at a 2-simplex $(ijk)$ $s$ (resp.~$1$) if we have $a_{ijk}=1$ (resp.~$a_{ijk}=0$). Then, we show that
\begin{align}
Z^{+}(01234)=e^{2\pi i q(a)}.
\end{align}
To see this, we utilize the hexagon equation for UMTC~\cite{Kitaevanyons}
\begin{align}
    \begin{split}
        R^{ac}_{e}F^{acb}_d R^{bc}_g&=F^{cab}_d R^{fc}_d F^{abc}_d,
    \end{split}
\end{align}
with $d=a\times b\times c, e=a\times c, f=a\times b$. Due to the invariance of the $F$ symbol under permutation of labels, it simplifies as
\begin{align}
    \begin{split}
        F^{abc}_d &= R^{ac}_{e}R^{bc}_g (R^{fc}_d)^{-1}.
    \end{split}
\end{align}
By putting the hexagon equations to~\eqref{eq:15jsimple} and noting that $F$ is real, and both $F$ and $R$ symbols are symmetric under permutation of labels, we can see that
\begin{align}
    \begin{split}
        Z^+(01234) &=   R^{012,234} (R^{014,234}R^{014,124}(R^{1234,014})^{-1}) ((R^{014,134})^{-1}(R^{014,123})^{-1}R^{1234,014})\\ &\times (R^{034,123}R^{034,013}(R^{0123,034})^{-1})((R^{034,023})^{-1}(R^{034,012})^{-1}R^{0123,034}) \\
        &= R^{012,234}\times (R^{014,234}R^{014,124}(R^{014,134})^{-1}(R^{014,123})^{-1}) \\ & \times (R^{034,123}R^{034,013}(R^{034,023})^{-1}(R^{034,012})^{-1}) \\
        &= \exp\left(2\pi i\cdot \frac{1}{4}(\hat{a}\cup \hat{a}+\hat{a}\cup_1\delta\hat{a})\right)=e^{2\pi i q(a) }.
    \end{split}
\end{align}
This equation shows that the Crane-Yetter path integral with a fixed anyon configuration specified by the 2-form field $a$ is identical to $z(\eta,a)$, since the normalization factor~\eqref{eq:normalizationFactor_15j} is unit for abelian UMTC. So if we further sum over the configuration $a\in Z^2(M,\mathbb{Z}_2)$, we obtain the partition function of the Crane-Yetter TQFT as
\begin{align}
\begin{split}
    Z_{\mathrm{CY}}(M)&=\mathcal{D}^{2(N_0-N_1)-\chi(M)}\sum_{a\in Z^2(M,\mathbb{Z}_2)}z(\eta=0,a) \\
    &=\sqrt{2}^{-\chi(M)}2^{(N_0-N_1)}|B^2(M,\Z_2)|\sum_{a\in H^2(M,\mathbb{Z}_2)}z(\eta=0,a) \\
    &= \sqrt{2}^{-\chi(M)}\frac{|H^0(M,\Z_2)|}{|H^1(M,\Z_2)|}\sum_{a\in H^2(M,\mathbb{Z}_2)}z(\eta=0,a) \\
    &=\frac{1}{\sqrt{|H^2(M,\mathbb{Z}_2)|}}\sum_{a\in H^2(M,\mathbb{Z}_2)}z(\eta=0,a)\\
    &=Z_{\mathrm{ABK}}(\eta=0).
    \end{split}
\end{align}
Thus, on oriented manifolds, the exotic invertible phase $Z_{\mathrm{ABK}}(\eta=0)$ is identical to the Crane-Yetter TQFT with the semion UMTC $\{1,s\}$ at the level of the microscopic action.
We note that the (3+1)d theory $z(\eta=0,a)=e^{2\pi i \int q(a) }$ in the oriented spacetime has been studied in~\cite{Tsui_2020} as a lattice model that realizes a (3+1)d SPT phase protected by 1-form $\Z_2$ symmetry. Also, a Hamiltonian model (the Walker-Wang model~\cite{WalkerWang2011}) for the Crane-Yetter TQFT based on the semion UMTC $\{1,s\}$ has been obtained in~\cite{von_Keyserlingk_2013}.

\section{Gu-Wen type phases}
\label{sec:guwen}
Finally, we propose an analogue of the Gu-Wen SPT phase based on the onsite (0-form) $G$ symmetry and the Wu structure of the spacetime, labeled by a pair of cohomological data
\begin{align}
    (\nu_d, n_{d-2})\in C_{\rho}^{d}(BG,U(1))\times Z^{d-2}(BG,\mathbb{Z}_2),
\end{align}
which are subject to the constraint similar to the Gu-Wen equation,
\begin{align}
    \delta_{\rho}\nu_{d}=\frac{1}{2}\mathrm{Sq}^3(n_{d-2}) \mod 1,
    \label{eq:guweneq}
\end{align}
where $\rho$ denotes the twisted $G$-action on $U(1)$ where the anti-unitary elements act by complex conjugation.

By utilizing the path integral of $z(\eta,a)$, it is a simple matter to consider the Gu-Wen type phase coupled with the Wu structure, for a given data of $(\nu_d, n_{d-2})\in C_{\rho}^{d}(BG,U(1))\times Z^{d-2}(BG,\mathbb{Z}_2)$ with  $\delta_{\rho}\nu_{d}=\frac{1}{2}\mathrm{Sq}^3(n_{d-2}) \mod 1$. For a given $G$-gauge field $g: M\to BG$, the action is defined as
\begin{align}
    Z(M;g,\eta)=z(\eta,g^*n_{d-2})\exp\left(2\pi i \int_{M}g^*\nu_{d}\right).
    \label{eq:guwenaction}
\end{align}
Due to the Gu-Wen type equation~\eqref{eq:guweneq}, one can see that the 't Hooft anomaly is canceled out in the expression of $Z(M;g,\eta)$, and therefore provides a topologically invariant theory.

\section{Discussion}
\label{sec:discussion}
In this paper, we studied a state sum path integral that realizes an exotic invertible topological phase based on Wu structure of the spacetime. In the study of the Gu-Wen type exotic $G$-SPT phases, we only considered the case of $G=G_0\times \Z_2^T$ with $\Z_2^T$ a time-reversal symmetry, and the total symmetry group is direct product of $G_0$ and the 1-form symmetry of the Wu structure. It should be interesting to consider the case where $G$ is not a direct product between unitary group and $\Z_2^T$, or the case of the nontrivial 2-group that involves $G$ and the 1-form $\Z_2$ symmetry of the Wu structure. 

For instance, one can think of the case $G=G_0\times \Z_2^T$ and $G_0$ has a 2-group structure with respect to the 1-form $\Z_2$ symmetry of the Wu structure, characterized by the Postnikov class $\omega\in H^3(BG_0,\Z_2)$. In that case, the Wu structure is twisted as $\delta\eta=w_1w_2+\omega$. Since $z(\eta,a)$ constructed in this paper has the dependence on $\eta$ in the form of $(-1)^{\eta\cup a}$, we expect the extra contribution to the 't Hooft anomaly given by the response action $(-1)^{\int \omega\cup a}$ in the twisted case. Hence, $z(\eta,a)$ for the twisted Wu structure is expected to have the 't Hooft anomaly characterized by the response action
\begin{align}
    (-1)^{\int \Sq^3(a)+\omega\cup a}.
\end{align}
Then, to define the Gu-Wen type phase using the pair $(\nu_d, n_{d-2})\in C_{\rho}^{d}(BG,U(1))\times Z^{d-2}(BG,\mathbb{Z}_2)$
in the form of~\eqref{eq:guwenaction}, we need a twisted version of the Gu-Wen equation
\begin{align}
    \delta_{\rho}\nu_{d}=\frac{1}{2}(\mathrm{Sq}^3(n_{d-2})+\omega\cup n_{d-2}) \mod 1.
\end{align}
It would be interesting to have an explicit lattice construction of these Gu-Wen type phases based on the twisted Wu structure.

While we focused on the Wu structure based on the 2-group $\delta\eta=w_1w_2$, it would also be interesting to consider the Wu structure realized by a higher $n$-group involving the Lorentz $O(d)$ symmetry and the $(n-2)$-form $\Z_2$ symmetry, given by the trivialization of the $(n+1)$-th Wu class $\delta\eta=\nu_{n+1}$~\cite{ManifoldAtlasWu, MS}. In general, the Wu class in $H^n(N,\Z_2)$ has a property that $\Sq^{n+1}(x)+\nu_{n+1}\cup x$ is exact on a closed $(d+1)$-manifold $N$, for any $x\in H^{d-n}(N,\Z_2)$. 

Hence, we expect that a generalization of $z(\eta,a)$ with $a\in Z^{d-n}(M,\Z_2)$ given in the form of $z(\eta,a)=z'(a)(-1)^{\int \eta\cup a}$ also exists for $n$-th Wu structure, where $z'(a)$ is regarded as a coboundary of a trivial cochain $\Sq^{n+1}(a)+\nu_n\cup a$ evaluated on a spacetime $d$-manifold $M$.
Since the $(-1)^{\int \eta\cup a}$ part has an 't Hooft anomaly characterized by a response action $(-1)^{\int\nu_{n+1}\cup a}$, we expect that the generalization of $z(\eta,a)$ for the $n$-group has an 't Hooft anomaly with the response action
\begin{align}
    (-1)^{\int \Sq^{n+1}(a)}.
\end{align}
It would be interesting to consider an explicit lattice construction of the theory $z(\eta,a)$ based on the higher Wu structure, and Kitaev or Gu-Wen type phases obtained by utilizing this theory.

\section*{Acknowledgements}
The author thanks Yu-An Chen and Srivata Tata for useful discussions and related collaborations.  The author is supported by the Japan Society for the Promotion of Science (JSPS) through Grant No.~19J20801. 

\appendix

\section{Review on higher cup product}
\label{app:cup}
A branching structure on a triangulation is a local ordering of vertices, which can be specified by an arrow on each 1-simplex $\braket{ij}$, such there are no closed loops on any 2-simplices. This defines a total ordering of vertices on every single $d$-simplex $\braket{0\dots d}$.
In this appendix, we review cochain-level product operation called higher cup product, whose definitions are based on branching structure of the triangulation. See also~\cite{tata2020geometrically} for a reference on higher cup product.

Let $M$ be a triangulated $d$-dimensional manifold.
Firstly, the cup product gives the product of cochains
\begin{align}
    -\cup-:C^k(M,\Z_n)\times C^l(M,\Z_n)\to C^{k+l}(M,\Z_n),
\end{align}
whose explicit form is written as
\begin{align}
    (a\cup b)(0,\dots, k+l) = a(0,\dots,k)b(k,\dots,k+l).
    \label{eq:cupdef}
\end{align}
Note that this definition of cup product depends on the branching structure on the triangulation, where the ordering of vertices on each $(k+l)$-simplex is specified as $0\to 1\to\dots \to k+l$.
The cup product satisfies the Leibniz rule at the cochain level,
\begin{align}
    \delta(a\cup b)=\delta a\cup b+(-1)^ka\cup\delta b.
\end{align}
According to the Leibniz rule, one can show that the cup product defines the product of cohomologies $H^k(M,\Z_n)\times H^l(M,\Z_n)\to H^{k+l}(M,\Z_n)$. Actually, for given $a\in Z^k(M,\Z_n)$, $b\in Z^l(M,\Z_n)$, the shift of these cocycles by coboundaries is evaluated as
\begin{align}
    (a+\delta A)\cup(b+\delta B)=a\cup b + \delta((-1)^k a\cup B+A\cup b+A\cup\delta B),
\end{align}
so this also shifts $a\cup b$ by a coboundary, thus defines a map between cohomologies. Such a product operation defined on cohomologies is called a cohomology operation.

As a generalization of the cup product, the higher cup product $\cup_{i}$ gives~\cite{Steenrod46}
\begin{align}
    -\cup_i-:C^k(M,\Z_n)\times C^l(M,\Z_n)\to C^{k+l-i}(M,\Z_n),
\end{align}
whose explicit form is written as
\begin{align}
    (a\cup_i b)(0,\dots,k+l-i)=\sum_{0\le j_0 <\dots<j_i\le k+l-i}(-1)^p \cdot a(0\to j_0,j_1\to j_2,\dots)b(j_0\to j_1,j_2\to j_3,\dots).
\end{align}
Here, the notation $i\to j$ denotes all vertices from $i$ to $j$, $\{i,i+1,\dots, i+j\}$. 
$p$ is the number of permutations need to bring the sequence of vertices
\begin{align}
    0\to j_0,j_1\to j_2,\dots , j_0+1\to j_1-1,j_2+1\to j_3-1,\dots
\end{align}
to the sequence
\begin{align}
    0\to k+l-i.
\end{align}
In particular, $\cup_0$ is identified as the cup product $\cup$ defined in~\eqref{eq:cupdef}.
The higher cup product is subject to the generalized Leibniz rule,
\begin{align}
    \delta(a\cup_i b)=(-1)^{k+l-i}a\cup_{i-1}b +(-1)^{kl+k+l}b\cup_{i-1}a+\delta a\cup_i b+(-1)^k a\cup_i\delta b,
\end{align}
which is regarded as that the non-commutative property of $\cup_{i-1}$ is controlled by the $\cup_{i}$ product.
According to the above Leibniz rule, for closed $a$ and $b$, one can see that $a\cup_i b$ is not necessarily closed, $ \delta(a\cup_i b)=a\cup_{i-1}b+(-1)^{kl+k+l}b\cup_{i-1}a$ for $a\in Z^k(M,\Z_n), \beta\in Z^l(M,\Z_n)$. Hence, the product $\cup_i$ doesn't give a cohomology operation for $i>0$. 

However, when we take $\Z_n$ as $\Z_2$ it turns out that the map
\begin{align}
\begin{split}
    \Sq^{d-i}(a)&: Z^k(M,\Z_2)\to Z^{k+d-i}(M,\Z_2) \\
    \Sq^{d-i}(a)&:=a\cup_{i+k-d}a
    \end{split}
\end{align}
 does give a cohomology operation. Actually, one can check that $\Sq^{d-i}(a+\delta A)=\Sq^{d-i}(a)+\delta(a\cup_{i+k-d} A+A\cup_{i+k-d}a+A\cup_{i+k-d-1}A+A\cup_{i+k-d}\delta A)$ by using the generalized Leibniz rule. This shows that $\Sq^{d-i}$ defines a map $H^k(M,\Z_2)\to H^{k+d-i}(M,\Z_2)$.
 
 For convenience, we extend the definition of the $\Sq^{d-i}$ operation to non-closed cochains. For a given $\lambda\in C^{k}(M,\Z_2)$, we define
 \begin{align}
     \Sq^{d-i}(\lambda)&: C^k(M,\Z_2)\to C^{k+d-i}(M,\Z_2) \\
    \Sq^{d-i}(\lambda)&:=\lambda\cup_{i+k-d}\lambda + \delta\lambda\cup_{i+k-d+1}\lambda.
 \end{align}
 One can immediately check that $\Sq^{d-i}$ commutes with the coboundary,
 \begin{align}
     \delta\Sq^{d-i}(\lambda) = \Sq^{d-i}(\delta\lambda).
 \end{align}

\section{The obstruction class for Wu structure}
\label{app:w1w2}
Wu structure is specified by a choice of a $(d-2)$-chain $E\in C_{d-2}(M,\mathbb{Z}_2)$ with $\partial E=S$, where $S\in Z_1(M^d,\mathbb{Z}_2)$ represents the Poincar\'e dual of the Wu class $w_1 w_2$. Here we explain how to prepare the chain $S$. We will see that $S$ is supported on the orientation-reversing wall $W$, and gives the Poincar\'e dual of the 2nd Stiefel-Whitney class $w_2$ of $W$.

We describe $w_2$ as a $(d-2)$-chain $W_2\in Z_{d-2}(M^d,\Z_2)$, and $w_1$ as a 1-cochain $w_1\in Z^1(M^d,\Z_2)$. Then $S$ is prepared by $S=W_2\cap w_1$. The cap product $\cap$ can be computed explicitly using the formula~\cite{Thorngrenthesis}
\begin{align}
    (0\dots j+k)\cap a=(j\dots k)\left(\int_{(0\dots j)}a\right)
    \label{eq:cap}
\end{align}
for $a\in C^j(M,\Z_2)$ and $(0\dots j+k)$ is a $(j+k)$-chain. Then one can extend it by linearity in adding up chains.

Once we take the barycentric subdivision of the triangulation of $M$, we can prepare $W_2$ as the set of all $(d-2)$-simplices of $M$, except for $(d-2)$-simplices supported on $W$. To see this, we note that the Poincar\'e dual of $w_2$ is given by the set of all 2-simplices. Meanwhile, since $W$ is oriented, the set of all $(d-2)$-simplices of $W$ that represents $w_1$ of $W$ gives a trivial $(d-2)$-chain, so we can take the above $W_2$ as the dual of $w_2$.

\begin{figure}[htb]
\centering
\includegraphics{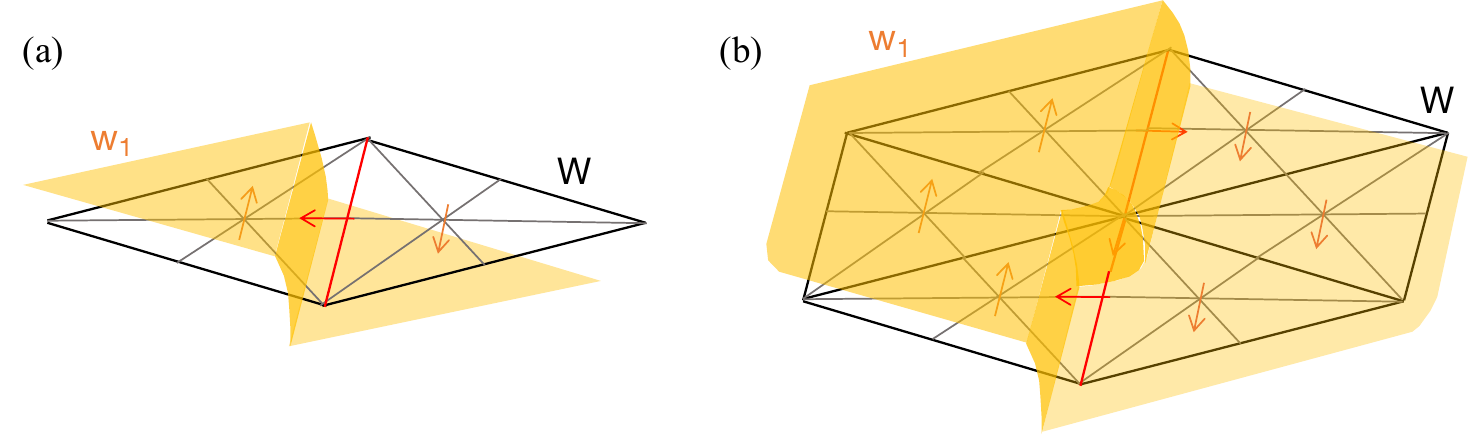}
\caption{(a): The figure of the orientation reversing wall for the spacetime dimension $d=3$. We perturb $W$ by a vector field perpendicular to $W$ (orange arrows) to obtain a perturbed object (yellow sheet). Then, on the zero locus (red line) we further perturb it by a vector field tangent to $W$ and normal to the zero locus (red arrow). (b): At the zero locus of the red vector field, e further perturb the yellow sheet to make the perturbed object intersects transversally with 1-simplices of $M^d$.}
\label{fig:perturbW}
\end{figure}

Then we explain how we prepare the cochain representative for $w_1$ of a $d$-manifold $M^d$. This is done by considering a perturbation of a codimension-1 orientation-reversing wall $W$ described as a yellow object in Fig.~\ref{fig:perturbW} for the case of $d=3$. That is, we think of shifting $W$ along a vector field expressed as orange vectors in Fig.~\ref{fig:perturbW}, which are perpendicular to the tangent of $W$. This vector field is thought of as a section of the normal bundle of $W$. Since the normal bundle of $W$ can be nontrivial, the vector field in general has a $(d-2)$-dimensional zero locus in $W$, shown as a red line in Fig.~\ref{fig:perturbW} (a). The zero locus is taken as a $(d-2)$-cycle of $W$.

At the $(d-2)$-dimensional zero locus, we further consider a vector field tangent to $W$ and perpendicular to the zero locus, shown as a red vector in Fig.~\ref{fig:perturbW} (a).
Then, we perturb the yellow object along the red vectors along the $(d-2)$-dimensional zero locus.
The red vector field is regarded as a section for the normal bundle of the zero locus in $W$, which again can have a $(d-3)$-dimensional zero locus, as shown in Fig.~\ref{fig:perturbW} (b). By repeating the process of shifting the yellow object along the $j$-dimensional zero locus to get a $(j-1)$-dimensional zero locus until we perform for $j=0$, we finally get a perturbation of $W$.

Then, the perturbation of $W$ intersects transversally with 1-simplices of $M$. Assigning the intersection number mod 2 on each 1-simplex defines a cocycle representative of $w_1$.

Now we compute the cycle representative of $w_1w_2$ using the representative of $w_1$ and $W_2$ defined above.
We take a branching structure of the barycentric-subdivided simplical complex, by labeling the barycenter of $i$-simplices by $(d-i)$.
According to the formula for the cap product~\eqref{eq:cap}, $S=W_2\cap w_1$ is given by a set of $(d-3)$-simplices $(j_0\dots j_{d-3})$, weighted by a number of $(d-2)$-simplices $(i,j_0\dots j_{d-3})$, where $w_1(ij_0)=1$ and $(i,j_0\dots j_{d-3})\in W_2$. 

\begin{figure}[htb]
\centering
\includegraphics{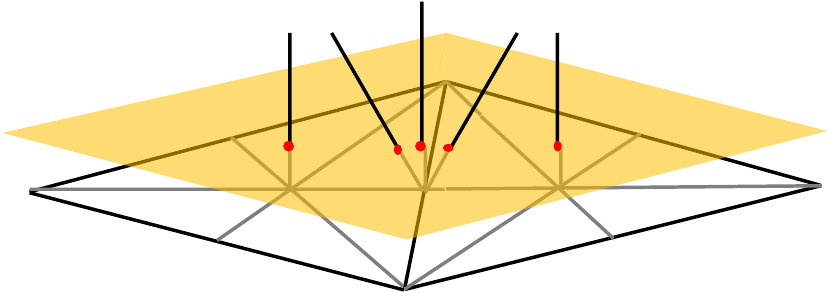}
\caption{The situation near the orientation reversing wall for the spacetime dimension $d=3$. The weight on each 0-simplex of $W$ corresponds to the intersection number between the perturbed sheet and $W_2$ near the 0-simplex. In this figure, the barycenter of a 2-simplex has the weight 1 while that of a 1-simplex has the weight 3.}
\label{fig:intersect}
\end{figure}

This $(d-3)$-chain $S$ can be rephrased as follows. $S$ is given by a set of $(d-3)$-simplices $(j_0\dots j_{d-3})$ of $W$ weighted by a number of $(d-2)$-simplices $(i,j_0\dots j_{d-3})$ whose vertex $i$ lies on a specific domain separated by $W$, and the domain is fixed by the perturbation of $W$ to define the cocycle $w_1$.
The weight on each $(d-3)$-simplex $(j_0\dots j_{d-3})$ on $W$ is regarded as the intersection number between $W_2$ and the $w_1$ sheet that occurs ``in the vicinity of'' $(j_0\dots j_{d-3})$, see Fig.~\ref{fig:intersect}.

\begin{figure}[htb]
\centering
\includegraphics{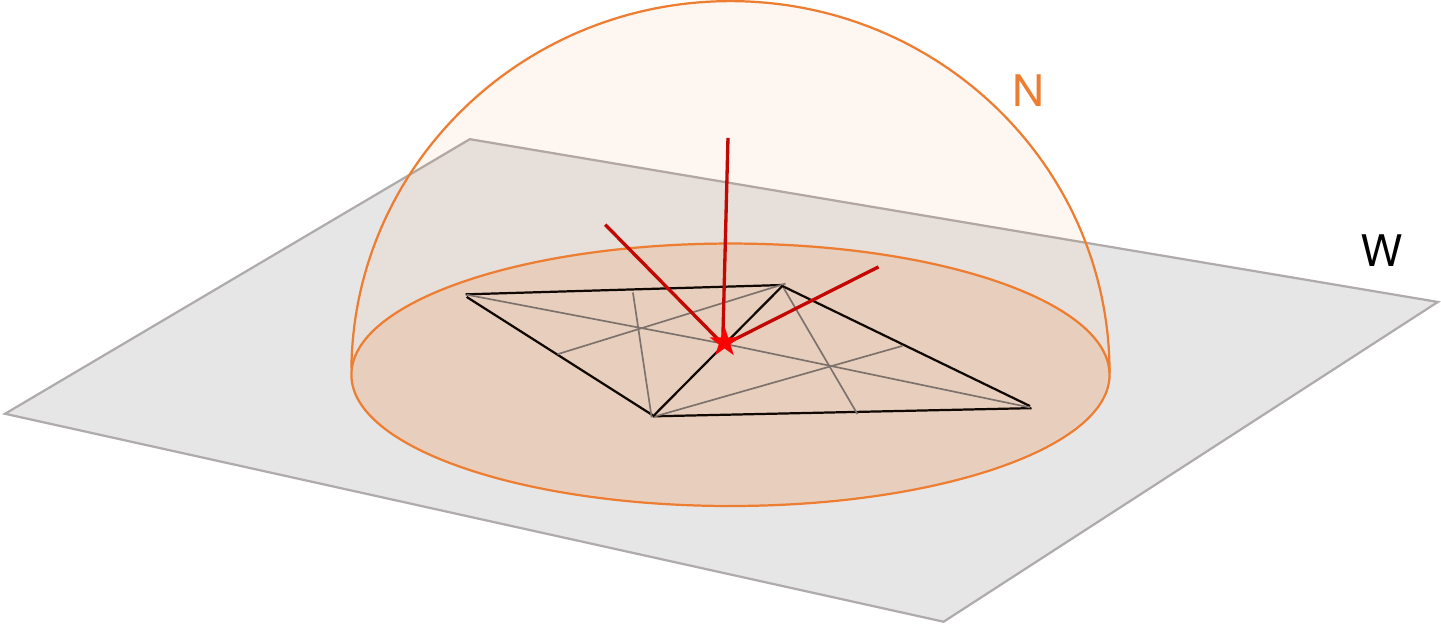}
\caption{The configuration of $N$ for $d=3$. Picking a 0-simplex of $W$ represented as a red star, then there are 1-simplices that evaluate nonzero $w_1$ around the 0-simplex, represented as red lines. $N$ contains all red 1-simplices in $N$ and $\partial N$ contains the starred vertex.}
\label{fig:Nconfig}
\end{figure}

Then, we can show that the weight for each $(d-3)$-simplex $W$ is always odd. 
To see that the weight is odd for a $(d-3)$-simplex $(j_0\dots j_{d-3})$, we take a non-closed $d$-submanifold $N$ as shown in Fig.~\ref{fig:Nconfig} such that $\partial N$ is oriented and contains $(j_0\dots j_{d-3})$, and $N$ contains all the $(d-2)$-simplices $(i,j_0\dots j_{d-3})$ counted in the weight. $N$ is contained in a specific domain of $M$ separated by $W$, and is taken to respect the original triangulation before taking barycentric subdivision, i.e., all the $(d-1)$-simplices on $\partial N$ are labeled as $(1,2,\dots d)$.

Let $S_{N}$ be a set of all $(d-2)$-simplices of $N$, except for those of $\partial N$. Let $S_{\partial N}$ be a set of all $(d-3)$-simplices of $\partial N$. We can then see that
\begin{align}
    \partial S_{N}=S_{\partial N},
\end{align}
which shows that the weight on $(j_0\dots j_{d-3})$ is odd, since $\partial S_{N}$ contributes to $(j_0\dots j_{d-3})$ from $(i,j_0\dots j_{d-3})$ in $S_{N}$ counted in the weight. Its derivation is essentially found in Appendix~\ref{app:bulkboundary}.
This proves that $S$ is the set of all $(d-3)$-simplices of $W$.

\section{Bulk-boundary Grassmann integral}
\label{app:bulkboundary}
In this appendix, we prove the following formula for the chains of barycentric subdivision frequently used in this paper. Let $N$ be a $d$-manifold with a non-empty boundary $\partial N$. Then we have 
\begin{align}
    \partial S_{N}=S_{\partial N},
\end{align}
where $S_N$ is a set of all $(d-2)$-simplices of $N$, and $S_{\partial N}$ is a set of all $(d-3)$-simplices of $\partial N$.

To see this, we introduce a theory for a bulk-boundary system $\sigma(\partial N,N;a,b)$ with $a\in C^{d-2}(\partial N,\Z_2)$ and $b\in Z^{d-1}(M,\Z_2)$, which satisfy $\delta a=b$ on the boundary $\partial N$.
$\sigma(\partial N,N;a,b)$ satisfies the following two properties:
\begin{enumerate}
    \item When $b$ is a coboundary $b=\delta\lambda$ and $a=\lambda$, we have 
    \begin{align}
        \sigma(\partial N, N; \lambda, \delta \lambda)=(-1)^{\int_{N}\Sq^2\lambda}(-1)^{\int_{S_N} \lambda}.
        \label{eq:sigmalambda}
    \end{align}
    \item When $b=0$ and $a$ is a coboundary $a=\delta\chi$, we have
    \begin{align}
        \sigma(\partial N, N; \delta\chi, 0)=(-1)^{\int_{\partial N}\Sq^2\chi}(-1)^{\int_{S_{\partial N}} \chi}.
        \label{eq:sigmachi}
    \end{align}
\end{enumerate}
Then, setting $\lambda=\delta\chi$ in~\eqref{eq:sigmalambda} gives
\begin{align}
        \sigma(\partial N, N; \delta\chi, 0)=(-1)^{\int_{\partial N}\Sq^2\chi}(-1)^{\int_{\partial S_{N}} \chi},
    \end{align}
    where we used $\delta\Sq^2(\chi)=\Sq^2(\delta\chi)$ reviewed in Appendix~\ref{app:cup}.
    Comparing this expression of $\sigma(\partial N, N; \delta\chi, 0)$ with~\eqref{eq:sigmachi} shows $\partial S_N=S_{\partial N}$.

The theory $\sigma(\partial N,N;a,b)$ with the above two properties is realized by a bulk-boundary version of the Grassmann integral constructed in~\cite{KOT2019, Kobayashi2019pin}. Since we deal with the oriented $N,\partial N$ in this paper, we present the proof only for the oriented case for simplicity. However, $\partial S_{N}=S_{\partial N}$ is also valid in unoriented cases.

Now let us write down the boundary Gu-Wen integral coupled with bulk;
we simply write the integral by $\sigma(a,b)$. 
We assign Grassmann variables $\theta_e, \overline{\theta}_e$ on each $(d-2)$-simplex $e$ of $\partial N$, and $\theta_f, \overline{\theta}_f$ on each $(d-1)$-simplex $f$ of $N\setminus M$.
We define the Gu-Wen integral as
\begin{equation}
    \sigma(a,b)=\int\prod_{f|b(f)=1}d\theta_f d\overline{\theta}_f \int\prod_{e|a(e)=1}d\theta_e d\overline{\theta}_e \prod_t u(t),
    \label{eq:GWboundary}
\end{equation}
$u(t)$ is a monomial of Grassmann variables defined on a $d$-simplex of $N$.  
$u(t)[b]$ is defined in the same fashion as in the case without boundary if $t$ is away from the boundary, which is introduced in Sec.~\ref{subsec:unoriented} of the main text. However, its definition gets modified when $t$ shares a $(d-1)$-simplex with the boundary. 
For simplicity, we assign an ordering on vertices of such $t=(01\dots d)$, so that the $(d-1)$-simplex shared with $M$ becomes $f_0=(12\dots d)$; the vertex $0$ is contained in $N\setminus M$. 
For instance, we can take a barycentric subdivision on $N$, and assign $0$ to vertices associated with $d$-simplices. 
We further define the sign of $(d-1)$-simplices on $M$, such that $f_0$ and $t$ have the same sign.

Then, $u(t)$ neighboring with $M$ is defined by replacing the position of $\vartheta_{f_0}$ in $u(t)[b]$ with the boundary action of the Grassmann integral on $f_0$, $u(f_0)[a]=\prod_{e\in f_0}\vartheta_e^{a(e)}$. 
We then have: On a $+$ simplex,
\begin{equation}
    u(t)=u(f_0)[a]\cdot\prod_{f\in\partial t, f\neq f_0}\vartheta_f^{b(f)}.
\end{equation}
On a $-$ simplex,
\begin{equation}
    u(t)=\prod_{f\in\partial t, f\neq f_0}\vartheta_f^{b(f)}\cdot u(f_0)[a].
\end{equation}
One can check that $u(t)$ defined above becomes Grassmann-even.
Then, one can see that the bulk-boundary Grassmann integral satisfies the quadratic property
\begin{align}
        \sigma(a+a', b+b')=\sigma(a,b)\sigma(a',b')(-1)^{\int_{\partial N} (a\cup_{d-3}a'+a\cup_{d-2}\delta a')+\int_N b\cup_{d-2}b'}.
        \label{eq:Quadboundarydisp}
    \end{align}
The proof for the quadratic property is found in~\cite{KOT2019}.

Now let us demonstrate~\eqref{eq:sigmalambda}. The quadratic part of $\sigma(\lambda,\delta\lambda)$ is determined by noting that the quadratic property
\begin{align}
   \sigma(\lambda+\lambda',\delta\lambda+\delta\lambda')= \sigma(\lambda,\delta\lambda)\sigma(\lambda',\delta\lambda')
   (-1)^{\int_{\partial N}\lambda\cup_{d-3}\lambda'}(-1)^{\int_N\lambda\cup_{d-3}\delta\lambda'+\delta\lambda'\cup_{d-3}\lambda},
\end{align}
is solved by $(-1)^{\int_N\Sq^2\lambda}$ up to a linear term. So, $\sigma(\lambda,\delta\lambda)$ can be expressed as
\begin{align}
    \sigma(\lambda;\delta\lambda)=(-1)^{\int_N\Sq^2\lambda}(-1)^{\sum_{e\in S'}\lambda(e)},
    \label{eq:formofsigmabbcoboundary}
\end{align}
with $S'$ some set of $(d-2)$-simplices $e$ of $N$. 
The linear term is fixed by computing $\sigma(\lambda;\delta\lambda)$ explicitly in the simplest case; $\lambda(e)=1$ on a single $(d-1)$-simplex, otherwise 0. 
Then we find that we get $\sigma(\lambda;\delta\lambda)=-1$ for arbitrary choice of a $(d-1)$-simplex $e$, which shows that $S'=S_N$.

One can also see~\eqref{eq:sigmachi} by noting that $\sigma(\partial N, N; \delta\chi, 0)$ reduces to the ordinary Grassmann integral $\sigma(\partial N; \delta\chi)$ supported solely on $\partial N$. Then,~\eqref{eq:sigmachi} is equivalent to~\eqref{eq:sigmacoboundary} in the main text by replacing $W$ with $\partial N$.

\bibliographystyle{ytphys}
\baselineskip=.95\baselineskip
\bibliography{ref}

\providecommand{\href}[2]{#2}\begingroup\raggedright\begin{thebibliography}{10}

\bibitem{Chen:2011pg}
X.~Chen, Z.-C. Gu, Z.-X. Liu, and X.-G. Wen, {\slshape {Symmetry Protected
  Topological Orders and the Group Cohomology of Their Symmetry Group},}
  \href{http://dx.doi.org/10.1103/PhysRevB.87.155114}{{\em Phys. Rev. B}
  {\bfseries 87} (2013) 155114},
\href{http://arxiv.org/abs/1106.4772}{{ arXiv:1106.4772~[cond-mat.str-el]}}.

\bibitem{Schnyder_2008}
A.~P. Schnyder, S.~Ryu, A.~Furusaki, and A.~W.~W. Ludwig, {\slshape
  Classification of topological insulators and superconductors in three spatial
  dimensions,} \href{http://dx.doi.org/10.1103/physrevb.78.195125}{{\em
  Physical Review B} {\bfseries 78} (Nov, 2008) },
  \href{http://arxiv.org/abs/0803.2786}{{ arXiv:0803.2786}}.

\bibitem{Kitaev2009free}
A.~Kitaev, {\slshape Periodic table for topological insulators and
  superconductors,} \href{http://dx.doi.org/10.1063/1.3149495}{{\em AIP
  Conference Proceedings} (2009) }, \href{http://arxiv.org/abs/0901.2686}{{
  arXiv:0901.2686}}.

\bibitem{FidkowskiKitaev2011}
L.~Fidkowski and A.~Kitaev, {\slshape {Topological phases of fermions in one
  dimension},} \href{http://dx.doi.org/10.1103/PhysRevB.83.075103}{{\em
  Physical Review B - Condensed Matter and Materials Physics} {\bfseries 83}
  (2011) 1--14}, \href{http://arxiv.org/abs/arXiv:1008.4138v2}{{
  arXiv:1008.4138v2}}.

\bibitem{Gu:2012ib}
Z.-C. Gu and X.-G. Wen, {\slshape {Symmetry-protected topological orders for
  interacting fermions: Fermionic topological nonlinear $\sigma$ models and a
  special group supercohomology theory},}
  \href{http://dx.doi.org/10.1103/PhysRevB.90.115141}{{\em Phys. Rev. B}
  {\bfseries 90} (2014) 115141},
\href{http://arxiv.org/abs/1201.2648}{{ arXiv:1201.2648~[cond-mat.str-el]}}.

\bibitem{Kapustin:2014dxa}
A.~Kapustin, R.~Thorngren, A.~Turzillo, and Z.~Wang, {\slshape {Fermionic
  Symmetry Protected Topological Phases and Cobordisms},}
  \href{http://dx.doi.org/10.1007/JHEP12(2015)052}{{\em JHEP} {\bfseries 12}
  (2015) 052},
\href{http://arxiv.org/abs/1406.7329}{{ arXiv:1406.7329~[cond-mat.str-el]}}.

\bibitem{Wang2017Interacting}
C.~Wang, C.~H. Lin, and Z.~C. Gu, {\slshape {Interacting fermionic
  symmetry-protected topological phases in two dimensions},}
  \href{http://dx.doi.org/10.1103/PhysRevB.95.195147}{{\em Physical Review B}
  {\bfseries 95} (2017) 1--32},
  \href{http://arxiv.org/abs/arXiv:1610.08478v1}{{ arXiv:1610.08478v1}}.

\bibitem{Witten2016Fermion}
E.~Witten, {\slshape {Fermion path integrals and topological phases},}
  \href{http://dx.doi.org/10.1103/RevModPhys.88.035001}{{\em Reviews of Modern
  Physics} {\bfseries 88} (2016) },
  \href{http://arxiv.org/abs/arXiv:1508.04715v2}{{ arXiv:1508.04715v2}}.

\bibitem{Metlitski2014Interaction}
M.~A. Metlitski, L.~Fidkowski, X.~Chen, and A.~Vishwanath, {\slshape
  {Interaction effects on 3D topological superconductors: surface topological
  order from vortex condensation, the 16 fold way and fermionic Kramers
  doublets},} \href{http://arxiv.org/abs/1406.3032}{{ arXiv:1406.3032}}.

\bibitem{Cheng2018Classification}
M.~Cheng, Z.~Bi, Y.~Z. You, and Z.~C. Gu, {\slshape {Classification of
  symmetry-protected phases for interacting fermions in two dimensions},}
  \href{http://dx.doi.org/10.1103/PhysRevB.97.205109}{{\em Physical Review B}
  {\bfseries 97} (2018) 1--11},
  \href{http://arxiv.org/abs/arXiv:1501.01313v3}{{ arXiv:1501.01313v3}}.

\bibitem{wanggu1703}
Q.-R. Wang and Z.-C. Gu, {\slshape {Towards a complete classification of
  fermionic symmetry protected topological phases in 3D and a general group
  supercohomology theory},}
  \href{http://dx.doi.org/10.1103/PhysRevX.8.011055}{{\em Physical Review X}
  {\bfseries 8} (2018) 011055}, \href{http://arxiv.org/abs/1703.10937}{{
  arXiv:1703.10937}}.

\bibitem{qingrui}
Q.-R. Wang and Z.-C. Gu, {\slshape Construction and classification of symmetry
  protected topological phases in interacting fermion systems,}
  \href{http://dx.doi.org/10.1103/PhysRevX.10.031055}{{\em Physical Review X}
  {\bfseries 10} (2018) 031055}, \href{http://arxiv.org/abs/1811.00536}{{
  arXiv:1811.00536}}.

\bibitem{Freed:2016rqq}
D.~S. Freed and M.~J. Hopkins, {\slshape {Reflection Positivity and Invertible
  Topological Phases},}
\href{http://arxiv.org/abs/1604.06527}{{ arXiv:1604.06527~[hep-th]}}.

\bibitem{Yonekura:2018ufj}
K.~Yonekura, {\slshape {On the Cobordism Classification of Symmetry Protected
  Topological Phases},}
  \href{http://dx.doi.org/10.1007/s00220-019-03439-y}{{\em Communications in
  Mathematical Physics} {\bfseries 368} (Apr, 2019) 1121–1173},
  \href{http://arxiv.org/abs/1803.10796}{{ arXiv:1803.10796~[hep-th]}}.

\bibitem{Guo:2018vij}
M.~Guo, K.~Ohmori, P.~Putrov, Z.~Wan, and J.~Wang, {\slshape {Fermionic
  Finite-Group Gauge Theories and Interacting Symmetric/Crystalline Orders via
  Cobordisms},} \href{http://dx.doi.org/10.1007/s00220-019-03671-6}{{\em
  Communications in Mathematical Physics} {\bfseries 376} (Jan, 2020)
  1073–1154}, \href{http://arxiv.org/abs/1812.11959}{{
  arXiv:1812.11959~[hep-th]}}.

\bibitem{Wan_2019}
Z.~Wan and J.~Wang, {\slshape Higher anomalies, higher symmetries, and
  cobordisms i: classification of higher-symmetry-protected topological states
  and their boundary fermionic/bosonic anomalies via a generalized cobordism
  theory,} \href{http://dx.doi.org/10.4310/AMSA.2019.v4.n2.a2}{{\em Annals of
  Mathematical Sciences and Applications} {\bfseries 4} (2019) 107–311},
  \href{http://arxiv.org/abs/1812.11967}{{ arXiv:1812.11967~[hep-th]}}.

\bibitem{Haah_2011}
J.~Haah, {\slshape Local stabilizer codes in three dimensions without string
  logical operators,} \href{http://dx.doi.org/10.1103/PhysRevA.83.042330}{{\em
  Physical Review A} {\bfseries 83} (Apr, 2011) },
  \href{http://arxiv.org/abs/1101.1962}{{ arXiv:1101.1962~[quant-ph]}}.

\bibitem{Yoshida_2013}
B.~Yoshida, {\slshape Exotic topological order in fractal spin liquids,}
  \href{http://dx.doi.org/10.1103/PhysRevB.88.125122}{{\em Physical Review B}
  {\bfseries 88} (Sep, 2013) }, \href{http://arxiv.org/abs/1302.6248}{{
  arXiv:1302.6248~[cond-mat.str-el]}}.

\bibitem{Vijay_2015}
S.~Vijay, J.~Haah, and L.~Fu, {\slshape A new kind of topological quantum
  order: A dimensional hierarchy of quasiparticles built from stationary
  excitations,} \href{http://dx.doi.org/10.1103/PhysRevB.92.235136}{{\em
  Physical Review B} {\bfseries 92} (Dec, 2015) },
  \href{http://arxiv.org/abs/1505.02576}{{
  arXiv:1505.02576~[cond-mat.str-el]}}.

\bibitem{Vijay_2016}
S.~Vijay, J.~Haah, and L.~Fu, {\slshape Fracton topological order, generalized
  lattice gauge theory, and duality,}
  \href{http://dx.doi.org/10.1103/PhysRevB.94.235157}{{\em Physical Review B}
  {\bfseries 94} (Dec, 2016) }, \href{http://arxiv.org/abs/1603.04442}{{
  arXiv:1603.04442~[cond-mat.str-el]}}.

\bibitem{Shirley_2019}
W.~Shirley, K.~Slagle, and X.~Chen, {\slshape Universal entanglement signatures
  of foliated fracton phases,}
  \href{http://dx.doi.org/10.21468/SciPostPhys.6.1.015}{{\em SciPost Physics}
  {\bfseries 6} (Jan, 2019) }, \href{http://arxiv.org/abs/1803.10426}{{
  arXiv:1803.10426~[cond-mat.str-el]}}.

\bibitem{hsin2021exotic}
P.-S. Hsin, W.~Ji, and C.-M. Jian, {\slshape Exotic invertible phases with
  higher-group symmetries,} \href{http://arxiv.org/abs/2105.09454}{{
  arXiv:2105.09454~[cond-mat.str-el]}}.

\bibitem{MS}
J.~W. Milnor and J.~D. Stasheff, {\em Characteristic classes}.
\newblock Princeton University Press, Princeton, N. J.; University of Tokyo
  Press, Tokyo, 1974.
\newblock Annals of Mathematics Studies, No. 76.

\bibitem{Gaiotto:2015zta}
D.~Gaiotto and A.~Kapustin, {\slshape {Spin TQFTs and Fermionic Phases of
  Matter},} \href{http://dx.doi.org/10.1142/S0217751X16450445}{{\em Int. J.
  Mod. Phys.} {\bfseries A31} (2016) 1645044},
\href{http://arxiv.org/abs/1505.05856}{{ arXiv:1505.05856~[cond-mat.str-el]}}.

\bibitem{Benini20192group}
F.~Benini, C.~Córdova, and P.-S. Hsin, {\slshape On 2-group global symmetries
  and their anomalies,} \href{http://dx.doi.org/10.1007/JHEP03(2019)118}{{\em
  Journal of High Energy Physics} {\bfseries 2019} (Mar, 2019) },
  \href{http://arxiv.org/abs/1803.09336}{{ arXiv:1803.09336~[hep-th]}}.

\bibitem{Cordova2019exploring}
C.~Córdova, T.~T. Dumitrescu, and k.~Intriligator, {\slshape Exploring 2-group
  global symmetries,} \href{http://dx.doi.org/10.1007/JHEP02(2019)184}{{\em
  Journal of High Energy Physics} {\bfseries 2019} (Feb, 2019) },
  \href{http://arxiv.org/abs/1802.04790}{{ arXiv:1802.04790~[hep-th]}}.

\bibitem{Tachikawa:2017gyf}
Y.~Tachikawa, {\slshape {On Gauging Finite Subgroups},}
  \href{http://dx.doi.org/10.21468/scipostphys.8.1.015}{{\em SciPost Physics}
  {\bfseries 8} (Jan, 2020) }, \href{http://arxiv.org/abs/1712.09542}{{
  arXiv:1712.09542~[hep-th]}}.

\bibitem{kapustin2015higher}
A.~Kapustin and R.~Thorngren, {\slshape Higher symmetry and gapped phases of
  gauge theories,} \href{http://arxiv.org/abs/1309.4721}{{
  arXiv:1309.4721~[hep-th]}}.

\bibitem{pontryaginsquare}
W.~Browder and E.~Thomas, {\slshape {Axioms for the generalized Pontryagin
  cohomology operations},}
  \href{https://academic.oup.com/qjmath/article-pdf/13/1/55/7288473/13-1-55.pdf}{{\em
  The Quarterly Journal of Mathematics} {\bfseries 13} (01, 1962) 55--60}.

\bibitem{Aharony2013reading}
O.~Aharony, N.~Seiberg, and Y.~Tachikawa, {\slshape Reading between the lines
  of four-dimensional gauge theories,}
  \href{http://dx.doi.org/10.1007/JHEP08(2013)115}{{\em Journal of High Energy
  Physics} {\bfseries 2013} (Aug, 2013) },
  \href{http://arxiv.org/abs/1305.0318}{{ arXiv:1305.0318~[hep-th]}}.

\bibitem{kapustin2013topological}
A.~Kapustin and R.~Thorngren, {\slshape Topological field theory on a lattice,
  discrete theta-angles and confinement,}
  \href{http://arxiv.org/abs/1308.2926}{{ arXiv:1308.2926~[hep-th]}}.

\bibitem{HalperinToledo}
S.~Halperin and D.~Toledo, {\slshape Stiefel-{W}hitney homology classes,}
  \href{http://dx.doi.org/10.2307/1970823}{{\em Ann. of Math. (2)} {\bfseries
  96} (1972) 511--525}.

\bibitem{BlantonMcCrory}
J.~D. Blanton and C.~McCrory, {\slshape An axiomatic proof of {S}tiefel's
  conjecture,} \href{http://dx.doi.org/10.2307/2042195}{{\em Proc. Amer. Math.
  Soc.} {\bfseries 77} (1979) 409--414}.

\bibitem{ManifoldAtlasWu}
{Karlheinz Knapp}, {\slshape Wu class,}
  \url{http://www.map.mpim-bonn.mpg.de/Wu_class}.

\bibitem{Kitaev00unpaired}
A.~Kitaev, {\slshape {Unpaired Majorana fermions in quantum wires},}
  \href{http://dx.doi.org/10.1070/1063-7869/44/10S/S29}{{\em Physics-Uspekhi}
  {\bfseries 44} (2001) 131}, \href{http://arxiv.org/abs/cond-mat/0010440}{{
  arXiv:cond-mat/0010440}}.

\bibitem{KirbyTaylor}
R.~Kirby and L.~Taylor, {\slshape {Pin Structure on Low-dimensional
  Manifolds},} \href{https://www3.nd.edu/~taylor/papers/PSKT.pdf}{{\em Geometry
  of low-dimensional manifolds} (1989) 177--242}.

\bibitem{Thorngren2018bosonization}
R.~Thorngren, {\slshape Anomalies and bosonization,}
  \href{http://arxiv.org/abs/arXiv:1810.04414}{{ arXiv:1810.04414}}.

\bibitem{Kobayashi2019pin}
R.~Kobayashi, {\slshape {Pin TQFT and Grassmann integral},}
  \href{http://dx.doi.org/10.1007/JHEP12(2019)014}{{\em Journal of High Energy
  Physics} {\bfseries 12} (2019) 014}, \href{http://arxiv.org/abs/1905.05902}{{
  arXiv:1905.05902}}.

\bibitem{crane1993}
L.~Crane and D.~Yetter, {\slshape {A categorical construction of 4D TQFTs},} in
  {\em Quantum Topology}, L.~Kauffman and R.~Baadhio, eds.
\newblock World Scientific, Singapore, 1993.
\newblock \href{http://arxiv.org/abs/arXiv:hep-th/9301062}{{
  arXiv:hep-th/9301062}}.

\bibitem{WalkerWang2011}
K.~Walker and Z.~Wang, {\slshape {(3+1)-TQFTs and topological insulators},}
  \href{http://arxiv.org/abs/arXiv:1104.2632v2}{{ arXiv:1104.2632v2}}.

\bibitem{Bhardwaj2020SL2Z}
L.~Bhardwaj, Y.~Lee, and Y.~Tachikawa, {\slshape ${SL}(2,\mathbb{Z})$ action on
  {QFT}s with $\mathbb{Z}_2$ symmetry and the {B}rown-{K}ervaire invariants,}
  \href{http://dx.doi.org/10.1007/JHEP11(2020)141}{{\em Journal of High Energy
  Physics} {\bfseries 2020} (Nov, 2020) },
  \href{http://arxiv.org/abs/2009.10099}{{ arXiv:2009.10099~[hep-th]}}.

\bibitem{morita}
S.~Morita, {\slshape On the pontrjagin square and the signature,}
  \href{https://repository.dl.itc.u-tokyo.ac.jp/records/39823#.YM3GUy33Lz8}{{\em
  Journal of the Faculty of Science, the University of Tokyo. Sect. 1 A,
  Mathematics} {\bfseries 18} (Dec, 1971) 405--414}.

\bibitem{Kitaevanyons}
A.~Kitaev, {\slshape {Anyons in an exactly solved model and beyond},}
  \href{http://dx.doi.org/10.1016/j.aop.2005.10.005}{{\em Annals of Physics}
  {\bfseries 321} (2006) 2--111},
  \href{http://arxiv.org/abs/cond-mat/0506438}{{ arXiv:cond-mat/0506438}}.

\bibitem{Bonderson07b}
P.~H. Bonderson, {\em Non-{A}belian {A}nyons and {I}nterferometry}.
\newblock PhD thesis, California Institute of Technology, 2007.

\bibitem{rowell2009classification}
E.~Rowell, R.~Stong, and Z.~Wang, {\slshape On classification of modular tensor
  categories,} \href{http://dx.doi.org/10.1007/s00220-009-0908-z}{{\em
  Communications in Mathematical Physics} (2009) 343--389},
  \href{http://arxiv.org/abs/0712.1377}{{ arXiv:0712.1377~[math.QA]}}.

\bibitem{Tsui_2020}
L.~Tsui and X.-G. Wen, {\slshape Lattice models that realize $\mathbb{Z}_n$-1
  symmetry-protected topological states for even $n$,}
  \href{http://dx.doi.org/10.1103/PhysRevB.101.035101}{{\em Physical Review B}
  {\bfseries 101} (Jan, 2020) }, \href{http://arxiv.org/abs/1908.02613}{{
  arXiv:1908.02613~[cond-mat.str-el]}}.

\bibitem{von_Keyserlingk_2013}
C.~W. von Keyserlingk, F.~J. Burnell, and S.~H. Simon, {\slshape
  Three-dimensional topological lattice models with surface anyons,}
  \href{http://dx.doi.org/10.1103/PhysRevB.87.045107}{{\em Physical Review B}
  {\bfseries 87} (Jan, 2013) }, \href{http://arxiv.org/abs/1208.5128}{{
  arXiv:1208.5128~[cond-mat.str-el]}}.

\bibitem{tata2020geometrically}
S.~Tata, {\slshape Geometrically interpreting higher cup products, and
  application to combinatorial pin structures,}
  \href{http://arxiv.org/abs/2008.10170}{{ arXiv:2008.10170~[hep-th]}}.

\bibitem{Steenrod46}
N.~Steenrod, {\slshape Products of cocycles and extensions of mappings,}
  \href{http://dx.doi.org/10.2307/1969172}{{\em Annals of Mathematics}
  {\bfseries 48} (1946) 290--320}.

\bibitem{Thorngrenthesis}
R.~Thorngren, {\slshape {Combinatorial Topology and Applications to Quantum
  Field Theory},} \href{https://escholarship.org/uc/item/7r44w49f}{{\em PhD
  Thesis} (2018) }.

\bibitem{KOT2019}
R.~Kobayashi, K.~Ohmori, and Y.~Tachikawa, {\slshape {On gapped boundaries for
  SPT phases beyond group cohomology},}
  \href{http://dx.doi.org/10.1007/jhep11(2019)131}{{\em Journal of High Energy
  Physics} (Nov, 2019) 1--2x}, \href{http://arxiv.org/abs/arXiv:1905.05391}{{
  arXiv:1905.05391}}.

\end{thebibliography}\endgroup

\end{document}